\RequirePackage{fix-cm}
\documentclass[twocolumn]{svjour3}          
\smartqed  
\usepackage{graphicx}
\usepackage{csquotes}

\usepackage{tabularx}
\usepackage{booktabs}

\usepackage{amsmath,amsfonts}
\usepackage{algorithmic}
\usepackage{algorithm}
\usepackage{array}
\usepackage[caption=false,font=normalsize,labelfont=sf,textfont=sf]{subfig}
\usepackage{textcomp}
\usepackage{stfloats}
\usepackage{url}
\usepackage{verbatim}
\usepackage{graphicx}
\usepackage{caption}
\usepackage{subcaption}
\usepackage{multirow}
\usepackage{booktabs}
\usepackage{array}
\usepackage[style=ieee]{biblatex}
\usepackage{hyperref}

\providecommand{\keywords}[1]
{
  \small	
  \textbf{\textit{Keywords---}} #1
}
\def\BibTeX{{\rm B\kern-.05em{\sc i\kern-.025em b}\kern-.08em
    T\kern-.1667em\lower.7ex\hbox{E}\kern-.125emX}}

\newcommand{\PS}{\mathit{PS}}

\bibliography{Main_manuscript}
\usepackage[usenames,dvipsnames]{color}

\begin{document}

\title{Quantifying Emotional Arousal through Pupillary Response: A Novel Approach for Isolating the Luminosity Effect and Predicting Affective States
}


\author{
Zeel~Pansara  \and
Gabriele~Navyte  \and
Tatiana~Freitas~Mendes  \and
Camila~Bottger  \and
Edoardo~Franco  \and
Luca~Citi  \and
Erik~S.~Jacobi  \and
Giulia~L.~Poerio  \and
Helge~Gillmeister  \and
Caterina~Cinel  \and
Vito~De~Feo
}


\institute{
Z. Pansara \and T. Freitas Mendes \and L. Citi \and C. Cinel \and V. De Feo \at
School of Computer Science and Electronic Engineering, University of Essex, Colchester, UK \\
\email{\url{zp20945@essex.ac.uk}, \url{vito.defeo@essex.ac.uk}}
\and
G. Navyte \at
ISER and Department of Psychology, University of Essex, Colchester, UK
\and
C. Bottger \and H. Gillmeister \at
Department of Psychology, University of Essex, Colchester, UK
\and
E. Franco \at
Politecnico di Torino, Torino, Italy
\and
E. S. Jacobi \at
Essex Business School, University of Essex, Colchester, UK
\and
G. L. Poerio \at
School of Psychology, University of Sussex, Brighton, UK
}

\date{Received: date / Accepted: date}

\maketitle

\begin{abstract}
Pupil dilation is recognized as an objective indicator of emotional arousal, but confounding factors such as the luminosity of stimuli and the surrounding environment have greatly limited its practical usefulness. This study presents a new approach to isolate and remove the effect of luminosity on pupil dilation. We validated this approach by showing 32 video clips with different content and emotional intensity to 47 participants, who reported their level of emotional arousal after each video. We developed a model capable of predicting the effect of luminosity on pupil size as a function of screen brightness, which adapts to individual physiological differences and different types of monitors through a brief pre-experimental calibration. We thus estimated the pupil size due exclusively to luminosity and subtracted it from the total recorded pupil size, obtaining the component due exclusively to arousal. From the latter, we predicted the arousal of each participant for each video using two models.
We first used a simple linear regression model. When we used the luminosity-corrected pupil size, we obtained a correlation between predicted and self-reported arousal of \(r\) = 0.65 $\pm$ 0.12, and \(R2\) of 0.43 $\pm$ 0.12. The uncorrected pupil size, instead, showed virtually no predictive power (\(r\) = 0.26 $\pm$ 0.15, \(R2\) = 0.09 $\pm$ 0.089).
We then used an Extreme Gradient Boosting model, obtaining even better results in the case of luminosity correction (\(r\) = 0.765 $\pm$ 0.047, \(R2\) = 0.556 $\pm$ 0.085). Our results highlight that separating emotional and luminosity components from pupillary responses is crucial for accurately predicting arousal. 

\keywords{AI (Artificial Intelligence) \and Pupil Dilation \and Pupil Size \and Emotion Detection \and Arousal \and Luminosity Effect \and Calibration \and Linear Regression (LR) \and Extreme Gradient Boosting (XGBoost)}
\end{abstract}

\section{Introduction}
\label{intro}

Emotions are a central component of human life, influencing most, if not all, aspects of cognition and behavior~\cite{izard2009emotion}. Arising from complex psychological and physiological processes, they shape how we perceive and interact with the world around us~\cite{gross2015emotion}, modulating key cognitive functions~\cite{tyng2017influences}, such as memory, attention, perception, and learning, and shaping our experiences~\cite{brosch2013impact}. Emotions also play an integral role in decision-making~\cite{lerner2015emotion} and in our social lives through their involvement in empathy, cooperation, facilitating meaningful interpersonal interactions~\cite{van2009emotions}, as well as in workplaces, as they can enhance team dynamics and leadership, contributing to better collaboration and overall productivity~\cite{kalraa2024role}. Recognizing one's own and others' emotions is crucial both in educational contexts~\cite{vistorte2024integrating} and in clinical settings, as it is vital for diagnosing and treating mental disorders~\cite{cooper2020facial}.

Emotional changes manifest through various physiological and behavioral indicators, such as heart rate, blood pressure, facial expression, body language, eye movements, and pupil dilation~\cite{shu2018review}. Each of these indicators provides valuable insights into emotional states~\cite{kassam2013effects}, supporting the development of comprehensive models of emotional and behavioral processes~\cite{zeidner2003development}. Accurately detecting emotional arousal has valuable applications in fields such as psychology~\cite{giannakakis2019review}, education~\cite{ba2023measuring}, human-computer interaction~\cite{peter2012emotion}, healthcare~\cite{dhuheir2021emotion}, and artificial intelligence~\cite{ballesteros2024facial}. AI-powered emotion detection systems have been proven helpful for optimizing emotion detection by analyzing facial expressions~\cite{ballesteros2024facial}, vocal tone~\cite{szelogowski2021emotion}, and physiological signals~\cite{wang2025emotion}, enabling more accurate and efficient identification of emotional states.

AI-powered systems using pupillary responses to measure emotional arousal have progressed more slowly than those using other physiological and behavioral markers due to the challenge of separating emotional arousal effects from luminosity effects, an issue we address in these studies~\cite{raiturkar2016decoupling, pan2024effects}. 
Foundational work by Bradley et al. established that while an initial light reflex occurs, the pupil's subsequent response is larger when viewing emotionally arousing stimuli, regardless of valence, a change mediated by sympathetic activity~\cite{bradley2008pupil}. Pupil dilation in response to emotional stimuli is indeed controlled by the autonomic nervous system (ANS) through the interplay of its two branches~\cite{lowenstein1950role, partala2003pupil}. The sympathetic nervous system (SNS) triggers pupil dilation (mydriasis) via the neurotransmitter norepinephrine, enhancing visual processing during arousal or stress~\cite{wu2022ocular}. The parasympathetic nervous system (PNS) regulates pupil constriction (miosis) by acting through the oculomotor nerve and the neurotransmitter acetylcholine. This function is most active during calm or restful states, reducing the amount of light entering the eye and adjusting pupil size for near vision~\cite{henderson2020autonomic}. Previous research shows that pupils often dilate within 200 milliseconds of the release of noradrenaline in response to emotional arousal~\cite{sirois2014pupillometry}. 

In addition to emotional arousal, ambient light levels also significantly influence pupil size.~\cite{pan2024effects}. Pupils typically constrict in bright conditions and dilate in darker environments~\cite{mathur2014influences}, with size varying by up to 50\%~\cite{spector1990pupils}, hindering researchers' use of pupil dilation as a direct measure of an individual's emotional state~\cite{cherng2020background}. A central challenge in pupillometry is, therefore, delineating the observed pupil size signal $PS(t)$ into its dominant Pupillary Light Response (PLR) term ($PS_{\text{luminosity}}$) and its psychologically relevant Psychosensory Pupil Response (PPR), or arousal term ($PS_{\text{arousal}}$). As Mathôt et al. and Strauch et al. detail, this challenge stems from the pupil acting as a \enquote{final common pathway} (or integrated readout) for both the light reflex and the emotional response~\cite{mathot2018pupillometry, strauch2022pupillometry}. Therefore, explicit modeling and removal of the luminance term is a necessary step for reliable arousal inference.  Establishing robust pre-processing and controlling for these luminance confounds are the \enquote{gold standards} for valid inference~\cite{kret2019preprocessing, mathot2022methods}. However, as the field moves toward more advanced dynamic modeling of pupil data~\cite{fink2023preprocessing}, it has become increasingly clear that traditional techniques like subtractive baseline correction are often insufficient for dynamic visual stimuli where luminance varies continuously.

Foundational work established the first mathematical models for predicting pupil size under controlled light. The unified formula of Watson and Yellott provides a parametric characterization of the steady-state PLR based on luminance, field size, and age~\cite{watson2012unified}. On the high-precision end of the engineering spectrum, Zhang et al. proposed a model based on Spatially Weighted Corneal Flux Density, a high-precision metric used to estimate pupil size by accounting for the non-uniform sensitivity of the retina to light~\cite{zhang2019pupil}. Extending this, Huang et al. showed that PLR is sensitive to the spatial and chromatic distributions of light across the retina, indicating that accurate modeling requires spatially\\ aware methods~\cite{huang2023pupil}. 

Nakayama et al. explored several mathematical formulations to express the relationship between pupil size and brightness, including power-log and exponential functions, before ultimately selecting a hyperbolic model to predict pupil size as a function of screen luminosity in a dark laboratory environment~\cite{nakayama2021controlling}. This model is personalized such that the specific coefficients of the hyperbolic function are determined experimentally for each individual to account for personal physiological differences in light adaptation. Although they used subjective evaluation scores and task conditions to validate the model's success in isolating signals related to mental activity, they did not utilize the resulting residual to predict participants' self-reported arousal.

An alternative research tradition treats the pupil as a dynamic system whose neural inputs are convolved with an impulse-response function. Wierda et al. used deconvolution to uncover fast attentional dynamics from pupil traces~\cite{wierda2012pupil}. This was refined by McCloy et al. who improved the alignment between model predictions and behavioral events by estimating the pupil impulse response~\cite{mccloy2016temporal}. Korn and Bach formalized this using the General Linear Model (GLM) to jointly estimate inputs from light and cognitive events~\cite{korn2016solid}. However, these studies use a temporal attention task with sequential, isoluminant, and discrete stimuli (letters) or, more generally, controlled cognitive events, i.e. stimuli in which the onset of the cognitive event is precisely known. In many laboratory conditions where naturalistic stimuli are used (e.g., emotionally charged videos), the arousal events and their timing are ambiguous, which violates the requirement for discrete and convoluted input signals. This framework was extended by Napieralski and Rynkiewicz to explicitly model the PLR for the purpose of decoupling light-driven dilation from other factors~\cite{napieralski2019modeling}. However, that study did not include validation where both factors-dynamic luminosity and emotional arousal are present simultaneously. In summary, the dynamic models presented in the above-mentioned studies are powerful for controlled, event-based experiments, but do not extrapolate to continuous naturalistic content like videos where luminance varies unpredictably. 

Asano et al. developed dynamic reaction models using layered neural networks to simulate pupil behavior during continuous temporal changes in brightness, which proved more effective than simple linear models for dynamic video viewing~\cite{asano2021neural}. Although this approach successfully reconstructed the light reflex to isolate signals related to mental activity induced by audio instructions, the authors did not utilize the model to predict participants' self-reported arousal. Recently, Fanourakis and Chanel improved a highly detailed biologically plausible dynamic model of the PLR, devised by Pamplona et al., using a numerical solver to attenuate the dynamic light response during continuous screen viewing~\cite{fanourakis2022attenuation, pamplona2009photorealistic}. However, they used this complex model in a generalized, group-level fitting procedure, i.e., the model's parameters (e.g., sphincter/dilator tension, time constants) were not determined by a separate, dedicated calibration task for each individual subject~\cite{fanourakis2022attenuation}. Recently, Cai et al. introduced Open-DPSM, a convolutional toolkit that models responses to luminance, contrast, and spatial weighting~\cite{cai2024open}. While Open-DPSM represents the state-of-the-art in sensory modeling, it requires specialized hardware (photometers) for physical luminance calibration and complex gaze-contingent setups. Furthermore, as a sensory-driven model, it has not been validated on emotional or affective stimuli.

For applications requiring continuous processing, data-driven signal-processing methods have been employed. Gao et al. used an Adaptive Interference Canceler (AIC) driven by a photodiode-measured ambient light signal to isolate an affective pupil response~\cite{gao2009monitoring}. Soleymani et al. provided a multimodal database and conducted emotion classification using standard statistical and spectral features (e.g., average, standard deviation, and Hippus effect) derived from the pupil signal, achieving a 63.5 \% classification rate for arousal (low, medium, high) using eye gaze features~\cite{soleyami2012multimodal}. This approach relied on standard filtering and feature selection but lacked an explicit computational model for dynamic PLR removal, reinforcing the need for targeted PLR mitigation. Further efforts to isolate affective signals in videos led to the \enquote{decoupling} framework proposed by Raiturkar et al.~\cite{raiturkar2016decoupling}. They utilized a linear model of the PLR to predict pupil size based on frame luminance, treating the residual between measured and predicted size as an indicator of emotional arousal. While this laid the groundwork for signal separation, the model's performance was primarily validated through emotional event peaks identified by independent coders rather than continuous participant self-reports. Later in this article, we will explain why participants' self-assessments allow us to build models with greater predictive power than the evaluation of stimuli by independent encoders. More recently, Tarnowski et al. utilized a subject-specific linear regression to remove the effect of movie luminance from the pupil signal before performing emotion classification~\cite{tarnowski2020eye}. Although this approach considered individualized modeling and achieved a high accuracy of 80\% on a 3-class classification task, it remains interesting to consider how a linear suppression model would perform when predicting continuous arousal. A possible limitation of this approach is that the model relies on a linear approximation for the inherently non-linear, exponential-like pupillary light reflex. 

A recent critical finding is that the interaction between arousal and light is multiplicative, not additive~\cite{pan2022arousal}. Pan et al. first demonstrated this for emotional arousal~\cite{pan2022arousal}. Crucially, Pan et al. later showed that emotional arousal interacts differently, with the strongest effects occurring at very low luminances ($<20 \text{ cd/m}^2$), confirming that arousal effects are not monolithic~\cite{pan2024effects}. These studies establish the non-linear structure required for accurate $PS_{\text{luminosity}}$ modeling before PPR can be isolated. Also, Cherng et al. demonstrated that background luminance levels significantly modulate affective responses, with greater dilation typically observed under darker conditions due to competing activations of the sympathetic and parasympathetic nervous systems~\cite{cherng2020background}. Conversely, the models that assume a linear relationship between arousal and luminosity presented in this introduction have a good predictive power, probably because under normal experimental conditions the variation in luminosity is small enough to allow for a linear approximation with limited error. However, it is clear that this interaction highlights the need for a calibration process capable of accurately mapping the non-linear relationship between light and pupil size for each individual and the use of non-linear models.

In conclusion, previous research confirms that PLR modeling is essential for arousal inference from pupil dilation, that the light-arousal interaction is non-linear, and that PLR suppression improves emotion inference. However, no prior computational method provides a single solution that simultaneously offers per-subject calibration, continuous frame-wise processing compatible with video, no need for discrete stimulus timing, and with direct validation of the arousal residual against self-report. The present study introduces such an approach by fitting an individual luminance-pupil function and subtracting its prediction to recover an arousal-sensitive residual that explains self-reported arousal substantially better than raw pupil size.
We recently established the foundation for this methodology in Pansara et al., where we demonstrated the technical feasibility of using a personalized non-linear model to isolate the $PS_{\text{luminosity}}$ component during video viewing~\cite{pansara2024towards}. However, that preliminary work was limited to the predictive accuracy of the light-effect model itself and did not include removing it to isolate emotional arousal effects. Furthermore, that study relied on a relatively small participant sample. The current work extends that effort by applying the model to a larger and more diverse sample, including participants recruited outside the university population, and shifting from signal isolation to the direct prediction of continuous affective states, providing the first real-time validation against participants' self-reports of arousal. 

In particular, in the present study, we extended and improved our previous work as follows:
\begin{enumerate}
    \item First, we tested the model we presented in~\cite{pansara2024towards} in a dark laboratory with 10 participants. 
    \item Second, we assessed the model's functionality and tested its robustness across varying luminosity levels in laboratory conditions. 
    \item Third, we calculated the portion of pupil size change due to emotional arousal of 47 participants watching 32 emotional video clips, by subtracting the portion of pupil size attributable to luminosity from the total pupil size.
    \item Finally, we implemented a machine learning framework using Gradient Boosting to validate the physiological relevance of our isolated residual. After removing the luminosity-driven component, we used the resulting arousal-sensitive residual to predict participants' self-reported emotional arousal.
    
\end{enumerate}

The primary goal of this study was to provide a realistic and scalable tool for researchers -- particularly in the field of psychology -- conducting laboratory experiments with naturalistic video stimuli, and without the need to use specialized hardware such as a photometer. Our goal was to address the ambient light conditions and dynamic screen luminance found in typical experimental settings, rather than addressing the general problem of pupil size variation with luminosity (e.g., with extreme luminosity variations in open environments). These are fundamental steps towards providing a tool to assess emotional arousal from pupil dilation in dynamic screen-based conditions, where luminosity levels can change almost constantly.  

\section{Methodology}
Our approach to developing a model for arousal detection consisted of three study phases:

\renewcommand{\labelenumi}{\Alph{enumi}.}
\begin{enumerate}
    
    \item Development of the Luminosity Effect Prediction Model (LEPM): In this study phase, we developed a model to predict pupil size changes purely caused by luminosity and remove its effect. 

    \item Development of the Arousal Detection Model (ADM): We trained a linear model to predict emotional arousal from participants' pupil size data while they watched emotional video clips. To do this, we considered two conditions: in the first, we corrected the measured pupil size by removing the effect of luminosity using the LEPM model; in the second, we left the signal uncorrected. We then trained a linear regression model in both conditions using the participants' self-reported arousal as the target variable. Finally, we compared the results in the two conditions. At this stage, we deliberately chose a simple linear regression model to maximize accessibility and ease of use for researchers, while maintaining high predictive performance.

    \item Development of the Gradient Boost Regression-based Self-reported Arousal Prediction Model (GBR-SAPM): To achieve even better performance (than the ADM) in predicting arousal from pupil dilation, we utilized an Extreme Gradient Boosting (XGBoost) regressor to predict continuous arousal states, employing statistical features extracted from both raw and luminosity-corrected pupil signals. To ensure robust, subject-independent generalization, the model was evaluated using a nested Leave-One-Participant-Out (LOPO) cross-validation framework. This approach allows the model to capture complex non-linear physiological dynamics while minimizing the risk of overfitting to individual participants.

\end{enumerate}

\subsection{Development of the Luminosity Effect Prediction Model} \label{Luminosity Effect Prediction Model Development}

Here, we refer to a monochrome image as one in which all pixels have one color only. In contrast, a non-monochrome image may contain pixels of different colors. Additionally, by primary colors, we refer to red, green, and blue (RGB) colors. For example, when referring to a primary-color monochrome image, we will indicate an image with all pixels of one color only: red, green, or blue. Note that images used in real-world applications are commonly non-monochrome and non-primary colors. Finally, by gray, we mean the color resulting from combining red, green, and blue equally, e.g., 65\% red, 65\% green, and 65\% blue. 

The LEPM development process involved several steps. First, we created a model that predicted the luminosity of a computer screen. Second, we developed a model that predicted pupil size in a dark laboratory using monochrome images (red, green, blue, and gray) with varying luminosities, which were tested on monochrome and non-monochrome emotionally neutral images. Third, we validated the model in normal laboratory conditions laboratory environment. 

\subsubsection{Computing the luminosity of a monochrome image on a computer screen in a dark laboratory}
The brightness of a monitor is a function of the RGB values of all pixels. In the case of a monochrome image, we can write:
\begin{equation}
L(r, g, b) = k \cdot f(r, g, b)
\label{eq1}
\end{equation}
where \( L \) is the luminosity value, \( k \) is a constant (scaling factor) that depends on the brightness/contrast settings of the monitor, and \((r, g, b) \in [0\%, 100\%]^3\) are the RGB intensity values of a monochrome image displayed on the screen~\cite{pansara2024towards}. The relationship between RGB intensity and luminosity values is non-linear for a commercial monitor. For example, considering a primary color, e.g., blue, the luminosity corresponding to an intensity value of 60\% is not twice as high as that corresponding to a value of 30\%. Instead, the increase in luminosity as a function of RGB intensity is smaller at lower intensity values and dramatically increases as the intensity approaches maximum~\cite{pansara2024towards}. Furthermore, as more colors are used, this non-linear relationship changes. Therefore, the function \(f\) in Equation \ref{eq1} cannot be easily described analytically; we tabulated its values in a look-up table to accurately calculate the luminosity of images or frames displayed on a computer screen. The input of our look-up table was an RGB value corresponding to three integers ranging from 0 to 100. Luminosity values were measured empirically. We created a set of images with different RGB values. The luminosity of each image was measured in a dark laboratory using an LX1010BS digital lux meter at a distance of 65 cm, following standard eye-tracking protocols to ensure controlled conditions for accurate gaze and pupil response measurements~\cite{pansara2024towards}. If we wanted to consider all values, we would have \(100^3\) entries and would have to generate \(100^3\) images. We sub-sampled this space to make the search computationally less expensive, producing 1330 images. When querying the table, the corresponding luminosity value was returned if the input matched a point. Otherwise, a weighted average of the eight nearest points was calculated, with weights inversely proportional to their distances. In the case of images composed of only one of the fundamental colors, instead of eight nearest neighbors, we used only two~\cite{pansara2024towards}.

We calculated the values in the lookup table using a Dell Precision M6500 24-inch monitor (0.34 steradians at 65 cm) set to 100\% brightness and 100\% contrast, which we will refer to as the “reference monitor”~\cite{pansara2024towards}.
Different monitor models have different luminosity capabilities. However, when testing five monitors of different sizes and brands, we consistently observed that the non-linear relationship described in equation \ref{eq1} remains substantially valid, except for the scale factor \( k \) which must be recalculated every time the brightness and contrast settings are changed (see calibration procedure below). In particular, once the appropriate value of \( k \) was found, the maximum error between the monitors was 3 lux~\cite{pansara2024towards}, which is tolerable for our purposes. This suggests that, regardless of the manufacturer and model of the monitor, the relationship between the brightness values and RGB intensity is highly consistent between different monitors. The calibration procedure described below allowed us to compensate for the differences between the possible monitor settings and between different monitors, provided that the monitor settings did not change during the experiment. For the reference monitor, the value of \( k \) was 1~\cite{pansara2024towards}.

\subsubsection{Computing the luminosity of a non-monochrome image on a computer screen in a dark laboratory}
When an image or video frame is not monochrome, one can calculate the luminosity of each pixel and average the values obtained. However, we observed that the luminosity of an image is similar to that of a single-color image, with the average RGB value of the original image calculated by averaging the RGB values of all its pixels. To validate this observation, we selected 100 images: 50 taken from the Internet and 50 generated by assigning a random RGB value to each pixel. We then generated the corresponding monochrome images using the average RGB value of each original image. The luminosity of the original images differed from the luminosity of the corresponding monochrome images by 1.5 lux on average, with a maximum difference of 3 lux, tolerable for our purpose. For computational reasons, using the corresponding monochrome images is preferable, as it is much faster than calculating the luminosity for each pixel by querying our look-up table and then averaging~\cite{pansara2024towards}. 

\subsubsection{Predicting pupil size due to luminosity at different wavelengths}
The second step consisted of developing a model to predict the pupil size of a participant looking at a monochrome image on the screen\footnote{Throughout this study, we used an average of right and left pupil size as pupil size data.}.
This model incorporated the function \( f \), defined in the previous step, which transforms RGB values into luminosity. This luminosity function then served as an input to another function responsible for mapping luminosity to pupil size. For the choice of this second function, we were inspired by the works mentioned in the introduction and in particular by the work of Pamplona et al.~\cite{pamplona2009photorealistic} and the unified formula of Watson and Yellott~\cite{watson2012unified}, readapted by Fanourakis and Chanel~\cite{fanourakis2022attenuation}. All these works present a non-linear decreasing function that can be well approximated by an exponential, as also confirmed by Zhang et al.~\cite{zhang2019pupil}. For each primary color plus gray, our model can be described using the following equations:
\begin{flalign}
&\PS_{\text{red}}(r) = a_{\text{red}} \cdot e^{-b_{\text{red}} \cdot {f(r,0,0)}} + c_{\text{red}} \cdot {f(r,0,0)} + d_{\text{red}} && \nonumber \\
&\PS_{\text{green}}(g) = a_{\text{green}} \cdot e^{-b_{\text{green}} \cdot {f(0,g,0)}} \\& + c_{\text{green}} \cdot {f(0,g,0)}+ d_{\text{green}} && \nonumber \\
&\PS_{\text{blue}}(b) = a_{\text{blue}} \cdot e^{-b_{\text{blue}} \cdot {f(0,0,b)}} + c_{\text{blue}} \cdot {f(0,0,b)} + d_{\text{blue}} && \nonumber \\
&\PS_{\text{gray}}(w) = a_{\text{gray}} \cdot e^{-b_{\text{gray}} \cdot {f(w,w,w)}} \\& + c_{\text{gray}} \cdot {f(w,w,w)} + d_{\text{gray}} &&
\label{eq2}
\end{flalign}
where \(\PS\) is the pupil size, \(f(r,g,b)\) is part of the luminosity function that gives the predicted luminosity (see Equation \ref{eq1}), and \(a\), \(b\), \(c\), and \(d\) are coefficients to be determined by fitting the model to the recorded data~\cite{pansara2024towards}. The parameters \(b\) and \(c\) also absorb the parameter \(k\) in the Equation \ref{eq1}. Our goal was to assess the pupil response either to one of the fundamental wavelengths (red, green, blue) or to their uniform combination (gray). 

\begin{figure}[htbp]
\centering
\includegraphics[width=0.5\textwidth]{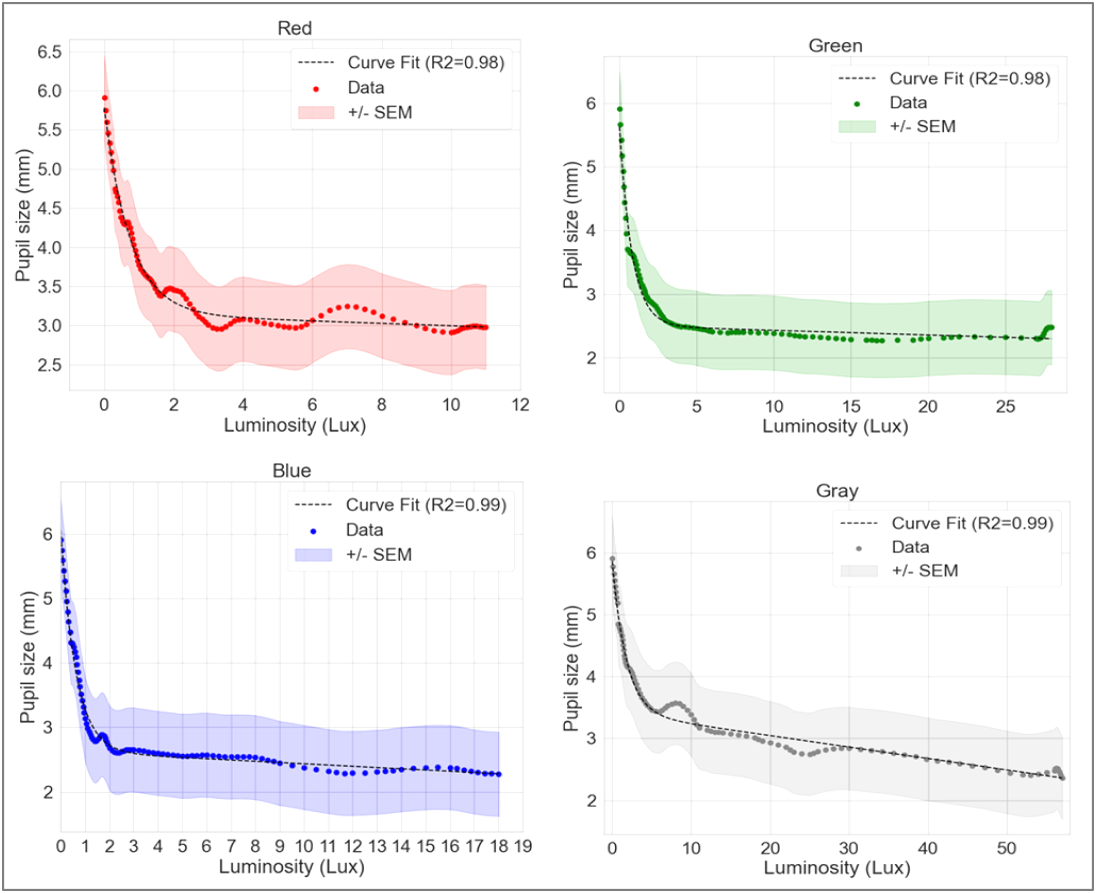}
\caption{Pupil size as a function of luminosity for red, green, blue, and gray
(scattered line for experimental data and dotted line for the fitted curve).}
\label{fig:all_model}
\end{figure}

To fit the model shown in Equation \ref{eq2}, for each of the four colors, we performed an initial estimation of the coefficients using data recorded from ten participants and then devised a method to recalibrate the coefficients \(a\), \(b\), \(c\), and \(d\) to new experimental participants. We sequentially presented 101 monochrome frames (video) of each of the four colors to the 10 participants for the initial estimation. We started from the lowest intensity (0\%, 0\%, 0\%) to the highest intensity (e.g., for red 100\%, 0\%, 0\% and for gray 100\%, 100\%, 100\%), increasing the brightness by 1\% every second. Each of the four videos lasted 101 seconds and was presented full-screen on the reference monitor. We measured pupil size for each video frame at a data rate of approximately\footnote{The data rate of the pupil size measured with the Tobii Pro Nano and iMotions software is variable.} 60 Hz and at a distance of 65 cm between the screen and the participant, using a Tobii Nano Pro eye tracker and the iMotions software (https://imotions.com/) in a dark (unlit) laboratory. At the same time, we measured luminosity during each video frame using the lux meter~\cite{pansara2024towards}. When an eye blink occurred (indicated by a value of -1 in the dataset output of the iMotions software), it was replaced with an estimated pupil size, which was calculated as the median pupil size for that specific frame. For each RGB intensity and participant, we considered the average pupil size recorded during the video frame presentation corresponding to that particular luminosity. Subsequently, we calculated the average pupil size data across all participants for each luminosity level. Figure \ref{fig:all_model} displays the average pupil size (aggregated across participants) plotted against luminosity for each color (scattered points). The dotted lines represent the fitted models\footnote{We did not need 101 measurements to fit a model with four coefficients, but using 101 data points allowed us to evaluate whether the model of the Equation \ref{eq2} was the most appropriate.}. As shown, pupil size diminished exponentially with rising luminosity~\cite{zhang2019pupil, pansara2024towards}. Figure \ref{fig:all_model} shows that the chosen model appropriately describes the data (\(R2\) $\ge$ 0.98).

\begin{table}[htbp]
\caption{\textit{Values of the four coefficients in Equation \ref{eq2}, for each color.}}
\centering
\begin{tabular}{|p{1cm}|p{1cm}|p{1cm}|p{1cm}|p{1cm}|}
\hline
Color& \textbf{\textit{\(a\)}} & \textbf{\textit{\(b\)}} & \textbf{\textit{\(c\)}} &  \textbf{\textit{\(d\)}}\\
\hline
Red & 2.6317 & 1.3371 & -0.0152 & 3.1500 \\
\hline
Green & 3.1259 & 1.2324 & -0.0073 & 2.5036 \\
\hline
Blue & 3.4430 & 1.6167 & -0.0193 &  2.6271 \\
\hline
Gray & 2.4465 & 0.5638 & -0.0184 & 3.4140 \\
\hline
\end{tabular}
\label{coeff_table}
\end{table}

Table \ref{coeff_table} shows the coefficient values of the model presented in Figure \ref{fig:all_model}~\cite{pansara2024towards}. These were then re-calibrated for each new participant. Indeed, it is essential to redetermine the coefficients before starting an experiment and tailoring them to the individual participant. To achieve this, each participant completed a calibration procedure before any experiment by watching a video of 27 monochrome frames~\cite{pansara2024towards}. The frames consisted of four colors --red, blue, green, and gray-- each displayed for 4 seconds, resulting in a total duration of 108 seconds. These images were generated using all possible combinations of three RGB intensity levels 0\%, 50\%, and 100\%, covering a range from the darkest (0\%, 0\%, 0\%) to the brightest (100\%, 100\%, 100\%).

Out of the 27 monochrome frames, nine were selected to calibrate the pupil size prediction model. These included three frames for each primary color (red, blue, and green) and gray. Since the black frame (0\%, 0\%, 0\%) is common to all colors, only one instance of it was used for calibration across all colors\footnote{In the case of red, the three calibration points were (0\%, 0\%, 0\%), (50\%, 0\%, 0\%), and (100\%, 0\%, 0\%), representing increasing intensity levels of red. Similarly, for green, the calibration points were (0\%, 0\%, 0\%), (0\%, 50\%, 0\%), and (0\%, 100\%, 0\%). For blue, they were (0\%, 0\%, 0\%), (0\%, 0\%, 50\%), and (0\%, 0\%, 100\%). For gray, the selected points were (0\%, 0\%, 0\%), (50\%, 50\%, 50\%), and (100\%, 100\%, 100\%). The remaining frames were used for testing.}.

Overall, the calibration process simultaneously accounts for individual variations in pupillary responses due to color, luminosity, and the difference between different monitors, establishing a reliable baseline for accurate subsequent measurements. 

\subsubsection{Two different approaches to predict the pupil size of a participant looking at a non-primary color monochrome image in a dark laboratory}
The model described by Equation \ref{eq2} enabled us to predict the pupil size of an individual observing a primary color or gray image with brightness levels ranging from 0 to 100. Our next objective was to extend the capability to predict pupil size in response to monochrome non-primary color images of any arbitrary color and luminosity, such as, for example, an image with RGB values (100, 75, 80), which represents a dark pink hue, with red as the dominant component. 

We achieved this with two approaches. The first approach, which we called “gray-based,” considered the pupil variation dependent only on the total luminosity of the image and not on the particular color. On the other hand, the second approach, called “color-based,” considered the luminosity of each primary color differently in the image. We expected that the first approach would work best for images where the three primary colors were fairly balanced, while the second approach would work best for images where one of the three primary colors was prevalent.

The “gray-based” approach involved calculating the image luminosity on the screen as a weighted average of the luminosities of the eight nearest images in the look-up table, as explained above. We then calculated the relative pupil size as if the image were equally composed of the three primary colors (according to the definition of gray used in this paper). In practice, we used the fourth equation of the set of Equations \ref{eq2}. To evaluate the “gray-based” approach and the calibration procedure, we tested 18 participants across five different monitors with varying resolutions, ranging from a minimum of 24 inches to a maximum of 49 inches (from 0.34 to 0.68 steradians). The test measurements included 18 (\(27-9 = 18\)) images from the calibration video and nine additional test images.

The “color-based” approach considered each image of a non-primary color \((r, g, b)\) as the superposition of three images of a single fundamental color: \((r, 0, 0)\), \((0, g, 0)\) and \((0, 0, b)\). This approach considers separately the effect of each wavelength on pupil size. For example, an image with RGB values \((64, 86, 45)\) was treated as three separate images: \((64, 0, 0)\), \((0, 86, 0)\) e \((0, 0, 45)\). We calculated the pupil size relative to each of the three images and assumed that the actual pupil size can be obtained as a linear combination of those three separate ones:
\begin{equation}
\begin{aligned}
\PS(r,g,b) &= \left(\frac{r}{r + g + b}\right) \cdot \PS_{\text{red}}(r) +\\ &+\left(\frac{g}{r + g + b}\right) \cdot  \PS_{\text{green}}(g)  + \\ &+\left(\frac{b}{r + g + b}\right) \cdot \PS_{\text{blue}}(b) 
\end{aligned}
\label{eq3}
\end{equation}
where \(\PS\) is predicted pupil size. The weights of each contribution to \(\PS\) in Equation \ref{eq3} are equal to the contribution of each fundamental color in the initial image~\cite{pansara2024towards}. For example, for a picture with RGB values \((64, 86, 45)\), we have: 
\begin{equation}
\begin{aligned}
\PS(64,86,45) &=\frac{64}{195} \cdot  \PS_{\text{red}}(64) + \\ &+\frac{86}{195} \cdot \PS_{\text{green
}}(86) + \frac{45}{195} \cdot \PS_{\text{blue}}(45).
\end{aligned}
\label{eq4}
\end{equation}
The “color-based” approach was tested simultaneously with the “gray-based” approach (with the same 18 participants, the same screens and resolutions).

\subsubsection{Combining the “gray-based” and the “color-based” approaches} 
\begin{figure}[!ht]
\centering
\includegraphics[width=0.5\textwidth]{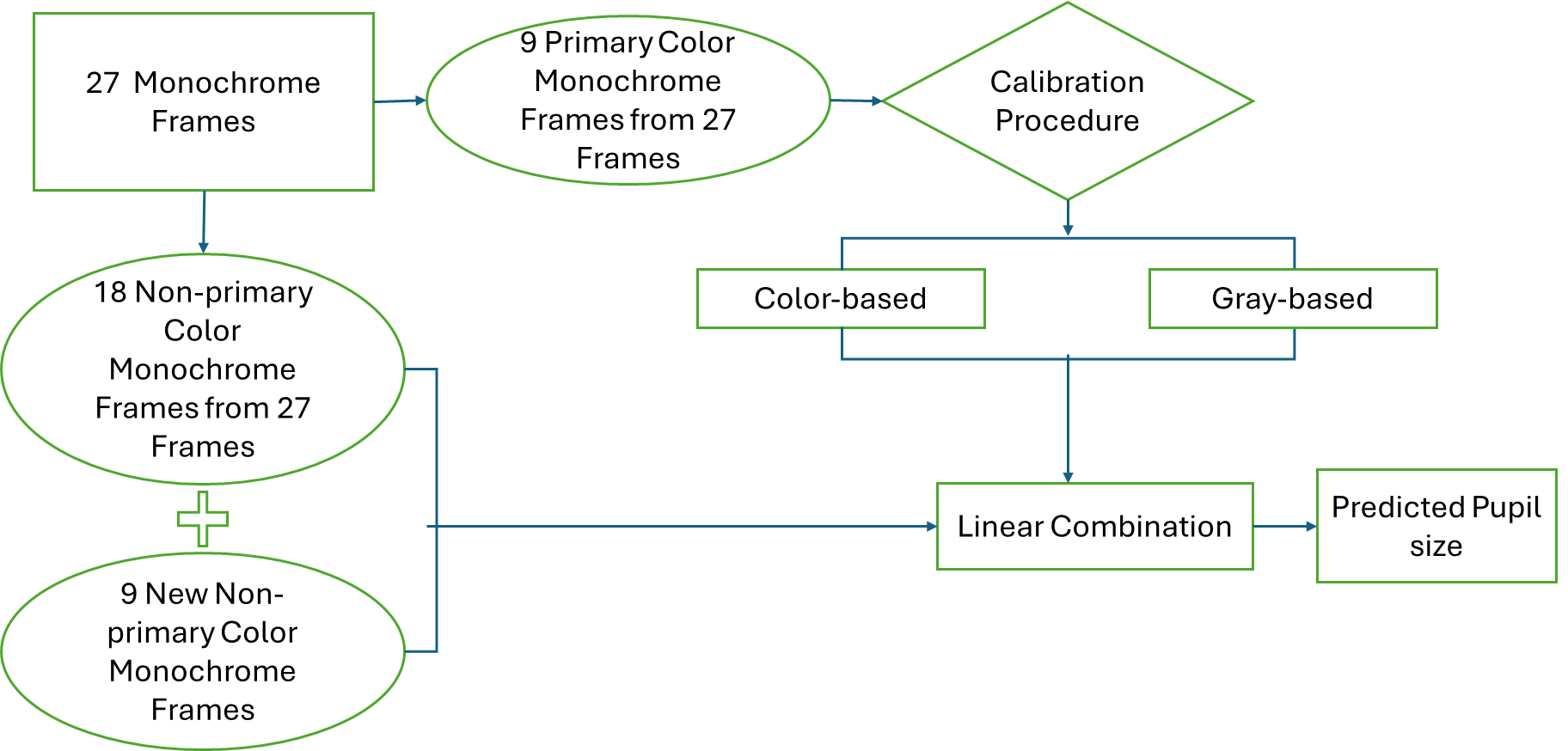}
\caption{Testing the combined approach.}
\label{LR_model}
\end{figure}

Since the “gray-based” method performed more accurately on images where the three primary colors had similar intensities,  while the “color-based” method excelled when one color was dominant, we combined both approaches to leverage their respective strengths. Consequently, the pupil size was calculated as a linear combination of the values obtained with the “gray-based” approach and those obtained with the “color-based” approach:
\begin{equation}
\begin{aligned}
\PS = K \cdot (a_\text{gray} \cdot \PS_\text{gray} + a_\text{red} \cdot \PS_\text{red} \\
+ a_\text{green} \cdot \PS_\text{green} + a_\text{blue} \cdot \PS_\text{blue}) + C \\
\end{aligned}
\label{regress_eq_ps}
\footnote{For simplicity sake, from now on we will omit that PS is a function of r, g, b.}
\end{equation}
with the constraint
\begin{equation}
\begin{aligned}
a_\text{gray} + a_\text{red} + a_\text{green} + a_\text{blue} = 1.
\end{aligned}
\label{constraint_eq_ps}
\end{equation}
We named this approach the \enquote{combined approach} (see Figure \ref{LR_model}). Given the constraint in Equation \ref{constraint_eq_ps}, it is expected that, after fitting, the value of the multiplicative coefficient \(K\) is close to 1 and that of the intercept \(C\) is close to 0. This would mean that the value of the pupil size \(\PS\) is given by the values  \(\PS_\text{gray}\), \(\PS_\text{red}\), \(\PS_\text{green}\), and \(\PS_\text{blue}\) added together in different percentages. 

The “combined” approach was tested simultaneously with the “gray-based” and the “color-based” approaches (with the same 18 participants, the same five different screens with different resolutions, the same images, etc.).  

For each participant, first, we calculated the values \(\PS_\text{gray}\), \(\PS_\text{red}\), \(\PS_\text{green}\), and \(\PS_\text{blue}\) using the nine images of the calibration procedure, as described above. Then, we trained and tested the model in Equation \ref{regress_eq_ps} (see Figure \ref{LR_model}).

Since the best results were obtained with the “combined” method, we used only this approach from then on and throughout this paper.

\subsubsection{Testing the Luminosity Effect Prediction Model in a normal laboratory conditions and with non-monochrome images}
\begin{figure}[!t]
\centering
\includegraphics[width=0.5\textwidth]{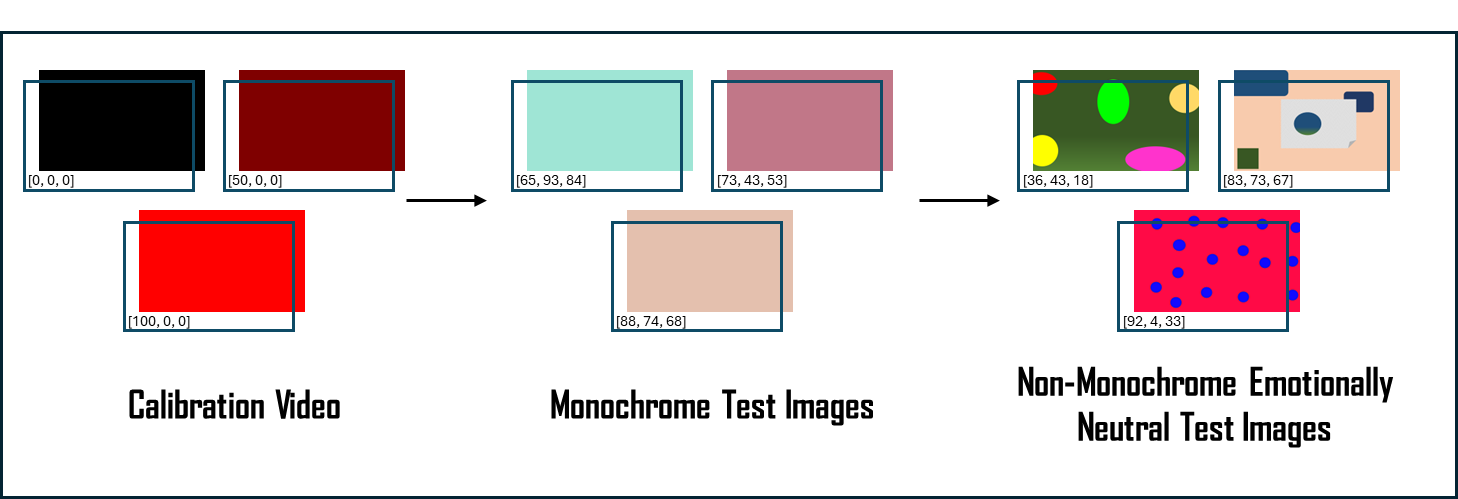}
\caption{Experiment flow of LEPM validation.}
\label{LPEM_experiment}
\end{figure}
Next, we tested our LEPM  model on non-monochrome images in a dark and a light environment. We recruited an additional 10 participants and used the 27 monochrome test images mentioned above, plus 46 non-monochrome emotionally neutral images (see Figure \ref{LPEM_experiment}).  Each image was displayed for 4 seconds, resulting in a video of 4 minutes and 52 seconds plus 36 seconds for the calibration procedure (9 images). We considered the mean pupil size across the 4 seconds for each image. Having taken the average value over a relatively long period compared to the time constant of the pupil size variation, we did not develop a dynamic model in this study. As before (when testing the model in a dark laboratory), we used five different monitors and computers/laptops with different brightness levels and resolutions.

The results obtained with the non-monochrome images were not as promising as those obtained with \\monochrome images. In particular, images that contained a very bright region within a dark figure yielded much worse results than monochrome images. This was because an image could be dark on average, but the participant looked at the bright region, potentially inducing a more pronounced pupillary constriction than that elicited by an image with its average luminosity level. Therefore, it was necessary to determine where participants were looking through the recorded eye-tracking data. Specifically, we analyzed a region with a 300-pixel radius centered on where the gaze was directed at each instant. If the average luminosity of that specific region exceeded that of the average luminosity of the entire image, we used that region average instead.  This method provided a more accurate representation of the subjective luminance and led to improved results, which we present in the Results Section \ref{results of LEPM}.

\subsection{Development of Arousal Detection Model } \label{Arousal detection Model (ADM) Development}
The second phase of the study aimed to demonstrate that the pupil size value discounted from luminosity was immediately usable for detecting emotional arousal levels without the need for complex pre-processing, advanced machine learning, or even deep learning models. 

We recruited 55 participants who, after the calibration procedure described above, watched 32 video clips with emotional content while pupil size was recorded. 
In order to facilitate a clear validation of the isolated pupil residual, the selected video clips were relatively short (with a duration variable between 10 and 100 second) and each eliciting a single primary emotional state (more details below).
After each video clip, participants self-reported the experienced emotions on a scale of 0 to 9 for 12 emotions (more details in Data Collection part \ref{data_collection}), which was used as the target value to predict self-reported arousal (more details in Arousal Ground Truth part \ref{GTA}). 

\subsubsection{Data Collection} \label{data_collection}

The categorization of emotions has long been debated, with two dominant approaches: basic emotion theory, which defines emotions as distinct categories, and multi-dimensional theory, which conceptualizes emotions as points along interconnected scales~\cite{zhang2020emotion}. This ongoing debate highlights the complexity of understanding human emotions. Our study adopted the multi-dimensional theory, utilizing Russell's widely recognized circumplex \\
model~\cite{russell1980circumplex}, which maps emotions along two key dimensions: valence (pleasantness) and arousal (activation level). Valence represents emotions ranging from unpleasant to pleasant, while arousal captures the intensity of emotional states, from calm to highly energized. These two dimensions are widely recognized in affective computing and cognitive research as a robust framework for modeling emotional experiences~\cite{kim2016identifying, ramaswamy2024multimodal} and have the advantage of providing a comprehensive yet efficient representation of complex emotional states. 

We selected 32 video clips with emotional content. These were classified into four distinct groups (8 videos per group) based on Russell's circumplex model~\cite{russell1980circumplex}: high arousal with positive valence, high arousal with negative valence, low arousal with positive valence, and low arousal with negative valence (this classification was tested through a pilot study with 37 participants). The duration of clips varied between 10 and 100 seconds, with different durations distributed uniformly across valence and arousal values and shown in a randomized order. 

As part of the participant recruitment process, we conducted a comprehensive screening procedure to identify potential indicators of conditions such as alexithymia, anxiety, depression, and personality disorders, which are known to influence emotional processing and responses~\cite{da2018alexithymia,gray2021emotion, kret2015emotion}. We initially screened 130 participants, with 55 meeting the inclusion criteria. Eligible participants were required to be between 18 and 65 years of age and to have sufficient English proficiency to provide informed consent and complete the questionnaires. Exclusion criteria included severe psychiatric conditions that could affect emotional recognition (e.g., schizophrenia, bipolar disorder, or autism spectrum disorder); however, mild anxiety, depression, or alexithymia were permitted, as assessed by the GAD-7, PHQ-9, and TAS-20. Additional exclusion criteria were intellectual disability, significant cognitive impairment, or recent neurological disorders (e.g., epilepsy, stroke, or traumatic brain injury within the past year). Eligibility was assessed using online Qualtrics survey questionnaires. The final sample enrolled in the study comprised 26 male and 29 female participants, aged 18 to 65 years (mean: 27.79 $\pm $10.360\footnote{In this paper, after the symbol $\pm$, we always indicate the standard deviation and never the standard error of the mean.}).

All participants provided informed consent before participation, and the study was approved by the University of Essex Ethics Committee (reference number: ETH2425-0176) as part of a more extensive study that included recording pupil size, facial expression data, galvanic skin response, photoplethysmogram, and electroencephalogram. Here, we only report on measurement and data processing related to pupil size.
The iMotions software (https://imotions.com/)~\cite{imotionsPupillometry101} managed the experimental protocol and facilitated efficient data acquisition.

\begin{figure}[!t]
\centering
\includegraphics[width=0.4\textwidth]{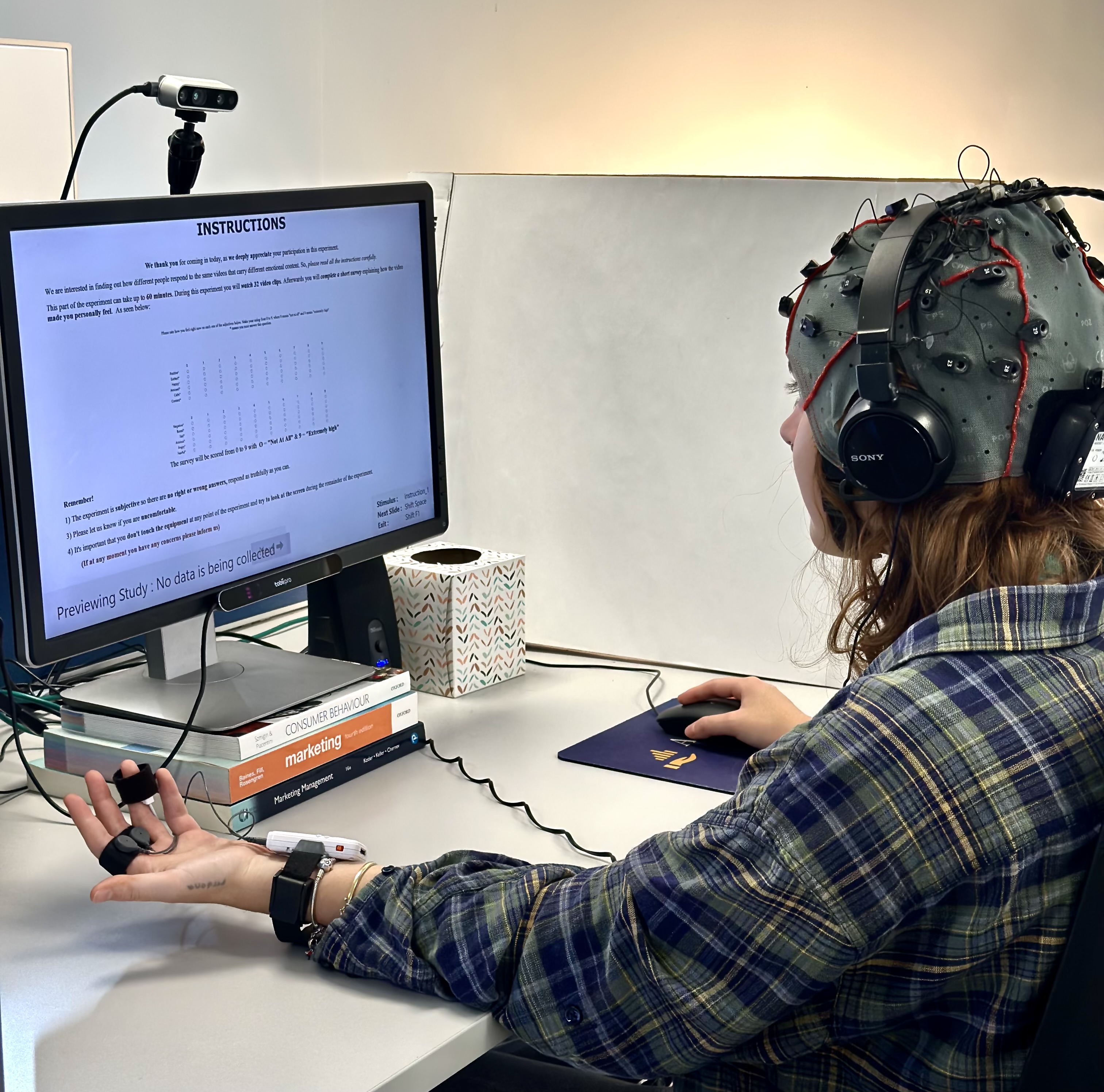}
\caption{Experimental setup.}
\label{experiment_setup}
\end{figure}

The experiment was conducted in a laboratory environment under normal lighting conditions, i.e., natural light during the day and in good weather, artificial light otherwise. (see Figure \ref{experiment_setup}). At the beginning of the experiment, and before displaying the video clips, participants were shown the calibration video described in section \ref{Luminosity Effect Prediction Model Development}. Participants were instructed to focus on the center of the screen during the calibration procedure.

Following calibration, the main experiment started. Each trial began with a 5-second gray screen displaying a central cross to neutralize previous emotional states, followed by an emotional video clip. To determine the emotional arousal, we only considered the salient intervals from each video (see below). After each clip, participants completed a questionnaire (extended from~\cite{kim2016identifying} to include a broader range of emotions) to assess 12 emotional states: positive, negative, happy, calm, content, amused, excited, angry, sad, disgusted, fearful, and bored. Participants rated their emotional responses on a scale from 0 (no emotion) to 9 (high emotion).

The final sample comprised 47 participants after excluding those based on three criteria: poor Tobii calibration that affected gaze and pupil size measurements, insufficient focus (e.g., participants consistently not looking at the screen or appearing to be asleep), and inconsistent responses, such as identical ratings for all questions. The excluded participants also showed low inter-rater reliability scores (Krippendorff's alpha $<$ 0.4) as their answers were highly inconsistent with the other participants' answers.
The pupil size data from these remaining participants were then pre-processed as explained in section \ref{Luminosity Effect Prediction Model Development}.

\subsubsection{Arousal Ground Truth} \label{GTA}

\begin{figure}[!t]
\centering
\includegraphics[width=0.55\textwidth]{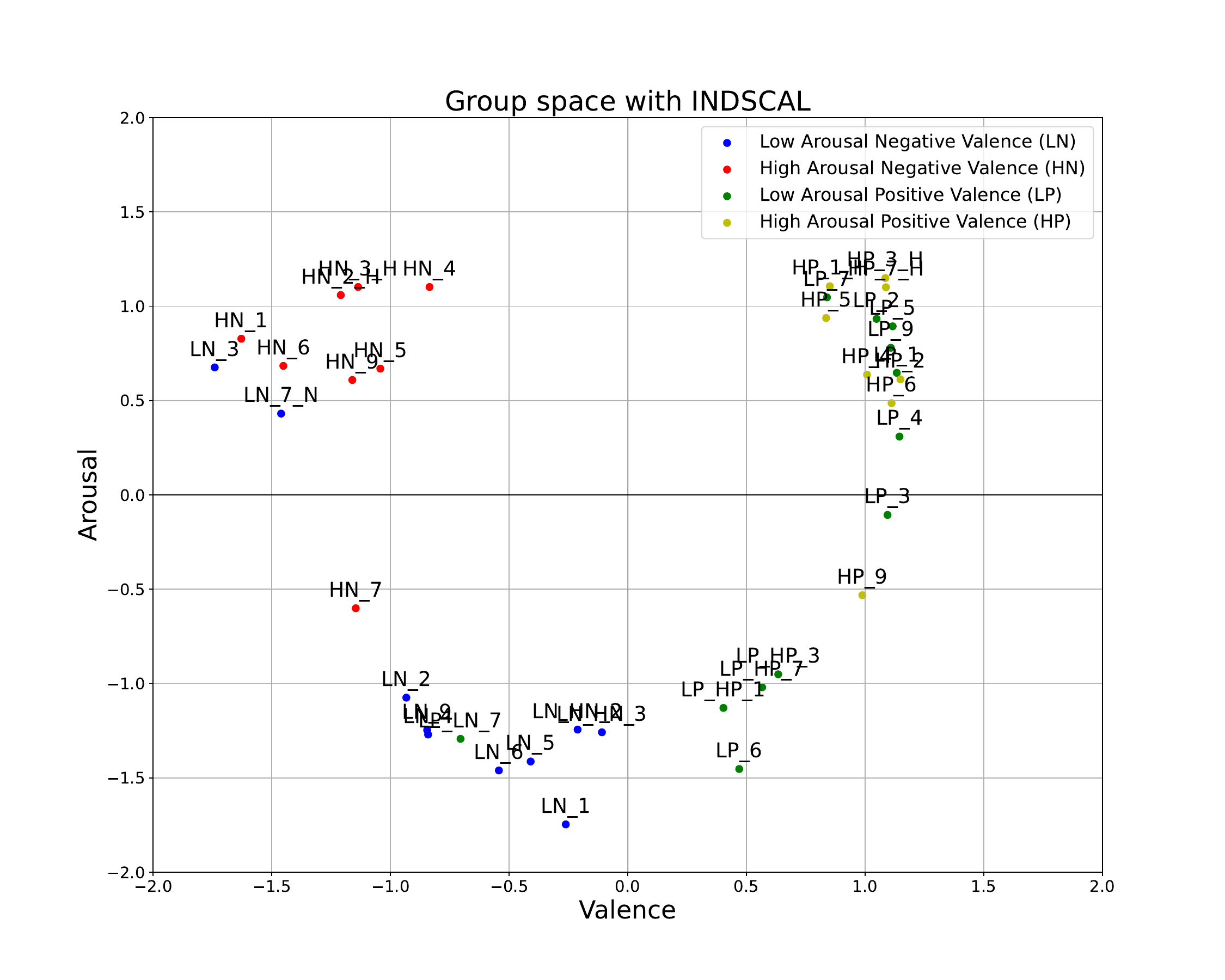}
\caption{Aggregate ground truth responses across all participants, categorized by stimuli arousal and valence: H = High Arousal, L = Low Arousal, P = Positive Valence, N = Negative Valence. Each point on the graph corresponds to a stimulus, i.e., a video clip. The valence and arousal values are rescaled in the range [-2, 2], where 0 indicates a neutral, average value.}
\label{group_space}
\end{figure}

As ground truth for the ADM, we used the participants' questionnaire responses, mapping their reported emotion scales into the arousal/valence two-dimensional space. Initially, we explored factor analysis for this purpose; however, this failed to distribute emotions effectively across valence and arousal dimensions, resulting in suboptimal model training outcomes (see results \ref{tab:indscal_vs_fa}). Instead, we employed Individual Scaling (INDSCAL) ~\cite{kim2016identifying}, a multidimensional scaling approach that reduces high-dimensional data into interpretable lower-dimensional representations. This approach condensed participants' responses from 12 emotional dimensions into two key dimensions: valence (x-coordinate) and arousal (y-coordinate) (see Figure \ref{group_space}). INDSCAL creates a common multidimensional space, like multidimensional scaling, and an individual space for each subject. This methodology improved the precision of predicting emotional states by providing a robust ground truth for training the ADM.

\subsubsection{Pupil Data Pre-processing}

The first pre-processing step involved identifying and marking blink-related artifacts. Data points identified as blinks by the eye tracker were substituted with null values. To account for the potential pre- and post-blink effects on pupil size measurements, an additional window of two milliseconds before and after each blink was also marked as null. Following this, we performed data imputation to fill the null values. The missing data points were replaced using a linear interpolation approach that considers the trends in the data immediately before and after the null segments. By interpolating in this way, we maximized the maintenance of the natural trajectory of pupil size changes over time, preserving the physiological relevance of the data while minimizing noise introduced by blinking.

\subsubsection{Arousal Effect Prediction}

After pre-processing the pupil size data, we applied the LEPM to predict the pupil size for each video frame across all video clips without influencing ambient and screen luminosity.

We trained the model described by the Equations \ref{regress_eq_ps} and \ref{constraint_eq_ps} by calculating \(\PS_\text{gray}\), \(\PS_\text{red}\), \(\PS_\text{green}\), and \(\PS_\text{blue}\) frame by frame and the coefficients \(a_\text{gray}\), \(a_\text{red}\), \(a_\text{green}\), \(a_\text{blue}\), \(K\), and \(C\) for each video clip. We hypothesized that since there was also an arousal component (in addition to the light component), the fitting would be worse and that the error in the fitting would be due specifically to the effect of arousal, given that the model was designed to capture only the component due to luminosity. In other words, we hypothesized that the measured pupil size would be equal to the pupil size predicted by the LEPM model as an effect of luminosity plus a residual:
\begin{equation}
\begin{aligned}
\PS_\text{measured} &= [K \cdot (a_\text{gray} \cdot \PS_\text{gray} + a_\text{red} \cdot \PS_\text{red}\\ 
&+ a_\text{green} \cdot \PS_\text{green} + a_\text{blue} \cdot \PS_\text{blue}) + C] \\& + Residual \\
\end{aligned}
\label{regress_eq_video}
\end{equation}
where 
\begin{equation}
\begin{aligned}
\PS_\text{luminosity} = [K \cdot (a_\text{gray} \cdot \PS_\text{gray} + a_\text{red} \cdot \PS_\text{red} \\
+ a_\text{green} \cdot \PS_\text{green} + a_\text{blue} \cdot \PS_\text{blue}) + C] \\
\end{aligned}
\label{regress_eq_video}
\end{equation}
so that
\begin{equation}
\begin{aligned}
\PS_\text{measured} = \PS_\text{luminosity} + Residual.
\end{aligned}
\label{regress_eq_video}
\end{equation}

The Residual was the portion of pupil size that cannot be explained by luminosity, i.e., the portion due to arousal:
\begin{equation}
\begin{aligned}
Residual = \PS_\text{arousal}.
\end{aligned}
\label{regress_eq_video}
\end{equation}
Hence, to extract arousal-related information for each video, we subtracted the predicted pupil size due to luminosity from the pre-processed measured pupil size:
\begin{equation}
\begin{aligned}
\PS_\text{arousal} = \PS_\text{measured} - \PS_\text{luminosity}.
\end{aligned}
\label{arousal_eq}
\end{equation}
The resulting difference represents the emotional arousal level of participants watching the video clips, as measured by pupil size.

\subsubsection{ADM Testing} 
To test the model, we compared its predictions of a participant's arousal while watching video clips - based on the measured pupil size - against the arousal ground truth. For each video and each participant, we had the self-reported arousal value (ground truth) \(\text{Arousal}_\text{self-reported}\), the recorded pupil size value \(\PS_\text{measured}\), and the pupil size value corrected for luminosity and due only to arousal \(\PS_\text{arousal}\).
We then calculated the Pearson correlation between the pupil size corrected and not corrected for luminosity and the self-reported arousal value (see results section).
One of the key challenges in this process was the considerable variation in the proportion of emotional content across different video clips. For example, some videos maintained a consistent emotional tone throughout their duration, while others contained emotional content only in specific segments, such as peaking toward the end while remaining neutral initially. 

When completing the questionnaire at the end of each video clip, participants were asked to report their emotion and arousal levels without specifying the exact moment in the video when they occurred. To identify the emotionally salient intervals, and  would allow us to predict arousal accurately at each relevant moment throughout the video duration, we randomly selected a sub-sample of 10 participants from the 55 who previously participated in the study. Participants were asked to indicate the salient intervals of each video. The final salient interval of a video was defined as the interval for which there was the highest degree of agreement among the 10 participants. The overlap between the chosen interval and that indicated by each participant was, on average, 74\% $\pm$ 15\%, considering all the videos, indicating a reasonable degree of agreement. For some videos, e.g., boring ones, the salient intervals corresponded to the entire video, while for other videos, e.g., scary ones, only some segments were salient. Finally, we only computed the average pupil size for each video and participant across the salient intervals, with and without our luminosity correction.

While our results demonstrated the potential of pupil size as a reliable indicator of emotional arousal, we wanted to demonstrate that \(\PS_\text{arousal}\), derived from our model, can be used without further processing or the use of complex machine learning techniques, becoming accessible to the entire scientific community, while still obtaining robust results. For this purpose we used the following procedure. We used a leave-one-participant-out cross-validation approach, which consisted of temporarily eliminating one participant at a time from our dataset as if they were a new participant. 

We then calculated the pupil size corrected for luminosity \(\PS_\text{arousal}\) and self-reported arousal\\ \(\text{Arousal}_\text{self-reported}\) for each video and each participant remaining in our dataset. We then assumed a simple linear relationship between the size of the pupil corrected for luminosity (component due to arousal) and self-reported arousal:
\begin{equation}
\begin{aligned}
\PS_\text{arousal} = a \cdot \text{Arousal}_\text{self-reported} + b
\end{aligned}
\label{arousal_ps_lin_eq}
\end{equation}

We fitted the model \ref{arousal_ps_lin_eq} to all the videos and all the participants (except one). We obtained a good fit (see Results). As mentioned in the introduction, it has recently been discovered that the interaction between the arousal effect and the effect of light on pupil size is multiplicative. However, we assumed, as a first approximation, that for relatively small variations in luminosity, a linear approximation could be used.

By inverting Equation \ref{arousal_ps_lin_eq}, we obtained an estimate of self-reported arousal \(\widehat{\text{Arousal}}_{\text{self-reported}}\) as a function of pupil size corrected for luminosity:
\begin{equation}
\begin{aligned}
\widehat{\text{Arousal}}_{\text{self-reported}} = \frac{\PS_{\text{arousal}} - b}{a}.
\end{aligned}
\label{ps_arousal_inverted_eq}
\end{equation}

We then repeated the same procedure for pupil size not corrected for luminosity. In this case, in Equations \ref{arousal_ps_lin_eq} and \ref{ps_arousal_inverted_eq}, \(\PS_\text{arousal}\) must be replaced with \(\PS_\text{measured}\). We predicted the self-reported arousal value for the eliminated participant starting from the pupil size, using Equation \ref{ps_arousal_inverted_eq}, and for all the videos. We compared the obtained values with the ground truth for the \\luminosity-corrected and uncorrected pupil size.  We repeated this procedure, eliminating one participant at a time, and finally calculated the average results for our sample.

\subsection{Development of the Gradient Boost Regression-based Self-reported Arousal Prediction Model (GBR-SAPM)}

We developed a relatively complex model for the first phase (LEPM) and a very simple one (ADM) for the second phase. As noted above, the ADM was deliberately designed to be simple in order to demonstrate that if a researcher simply subtracts the component $PS_{\text{luminosity}}$, obtained through the LEPM, from the measured pupil size, they obtain a significant improvement in assessing the experienced emotional arousal-- without resorting to further pre-processing or machine learning techniques (see Results). 

Nevertheless, when more advanced pre-processing and machine learning methods are applied, predictive performance improves further, as we demonstrate in this section. 

We first calculated the \(\PS_\text{arousal}\) signal, using the LEPM model, as described in Equation \ref{arousal_eq}. Then, from each of the three signals \(\PS_\text{measured}\), \(\PS_\text{luminosity}\), and \(\PS_\text{arousal}\), we extracted the following features:

\begin{itemize}
    \item Statistical Moments: Mean, variance, skewness, and kurtosis of the pupil diameter. These metrics provide a summary of the overall activation level and the distribution of the pupil response over the stimulus duration~\cite{bradley2008pupil}.
    \item Dynamic Features: The first and second derivatives of the pupil response. These features capture the temporal dynamics of dilation, which are critical for distinguishing the slower, sympathetic-driven arousal response from faster, reflex-driven changes~\cite{cacioppo2007handbook}.
\end{itemize}

The decision to extract features from all three of the above-mentioned signals, rather than using \(\PS_\text{arousal}\) only, stems from the recent discovery that the relationship between total pupil size, the part driven by emotional arousal, and the part driven by luminosity, is not additive but involves non-linear interactions\cite{pan2022arousal}.

We then used a Gradient Boosting regressor (XGBoost), which was preferred over traditional linear models due to its ability to model complex, non-linear interactions. XGBoost is particularly robust against multicollinearity, a common problem in physiological signal processing, and uses regularization (L1/L2) to prevent overfitting.

To ensure the model generalizes to unseen subjects, we implemented a nested Leave-One-Participant-Out (LOPO) cross-validation framework:
\begin{enumerate}
    \item Outer Loop (Generalization): The model is trained on $N-1$ participants and tested on the held-out participant. This mimics a real-world scenario where the system encounters a new user without prior calibration~\cite{leinonen2024empirical}.
    \item Inner Loop (Optimization): Inside each training set, we performed a five-fold cross-validation to optimize hyper parameters (e.g., learning rate, max depth).
\end{enumerate}
We evaluated the model's ability to track subjective arousal using the Coefficient of Determination ($R2$), Pearson correlation coefficient ($r$), and Normalized Root Mean Square Error ($NRMSE$).

We also trained the model to predict valence, although we expected worse results than for arousal. While there is consensus in the literature that pupil size is a reliable indicator of arousal, its relationship with valence is much more controversial. There are some studies, however, such as that by Tarnowski et al.~\cite{tarnowski2020eye}, in which machine learning (SVM and neural networks) was used to classify six basic emotions (happiness, sadness, fear, etc.), which inherently implies the distinction between positive and negative valence.
Inspired by this study, we trained our model to also predict emotional valence, and the training was done completely independently from the training done to predict emotional arousal.

All the code relative to the models presented in this paper was written in Python with libraries like pandas, numpy, scikit-learn, matplotlib, seaborn, sklearn, xgboost, and os.

\section{Results}

\subsection{Results of the Luminosity Effect Prediction Model (LEPM)} \label{results of LEPM}

The model illustrated in section “A. Development of Luminosity Effect Prediction Model” effectively predicts pupil size based on the RGB intensity values of the images on the screen in both dark and bright environments. The calibration procedure described above ensures that the method is flexible and adapted to inter-subjective differences, as well as the settings and type of screen used. 
As explained in section \ref{Luminosity Effect Prediction Model Development}, we initially used two different approaches, named “gray-based” and the “color-based”, and then we used a method consisting of combining both approaches, named the “combined”  approach. We assessed each approach using leave-one-image-out cross-validation, i.e., training the model on 27 images, eliminating one image at a time, and predicting the pupil size measured when the eliminated image was shown.

The results obtained from the 18 participants tested in dark laboratory conditions -- mean Pearson correlation coefficient with mean and max p-value, mean \(R2\),  mean \(NRMSE\)\footnote{Normalized Root Mean Squared Error (\(NRMSE\)) is a scale-independent error metric that allows comparison between datasets or models with different scales. We computed it by dividing the Root Mean Squared Error (RMSE) by a normalization factor, consisting of the range (max-min) of the observed data.} and mean percentage error -- are shown in Table \ref{comparison_per_part_results}. 

Table \ref{comparison_results}, on the other hand, shows the results obtained by combining the data from all 18 participants as if they were a single participant.

\begin{table}[htbp]
\caption{\textit{Results of all methods on monochrome images in a dark laboratory - average across participants.}}
\centering
\begin{tabular}{|p{1.6cm}|p{1.7cm}|p{1cm}|p{1.2cm}|p{1.4cm}|}
\toprule
\textbf{Method} & \textbf{Correlation} & \textbf{\(R2\) Score} & \textbf{\(NRMSE\)} &  \textbf{Average Error}\\
\midrule
Color-Based & 0.62 $\pm$ 0.124 (mean p =  0.0038, max p =  0.0214) & 0.40 $\pm$ 0.153  & 0.23 $\pm$ 0.048 & 6.52 $\pm$ 0.221 \% \\
\midrule
Gray-Based & 0.72 $\pm$ 0.150 (mean p =  0.0056, max p =  0.086) & 0.55 $\pm$ 0.193 & 0.21 $\pm$ 0.055 & 8.76 $\pm$ 0.290 \% \\
\midrule
Combination of color and gray-based approach & 0.84 $\pm$ 0.070 (mean p =  $<$ \(10^{-7}\), max p =  0.00001) & 0.70 $\pm$ 0.114 & 0.13 $\pm$ 0.028 & 6.75  $\pm$ 0.266 \%\\
\bottomrule
\end{tabular}
\label{comparison_per_part_results}
\end{table}

\begin{table}[!htbp]
\caption{\textit{Results of all methods on monochrome images in a dark laboratory - aggregating the data from all the participants}}
\centering
\begin{tabular}{|p{1.6cm}|p{1.7cm}|p{1cm}|p{1.2cm}|p{1.4cm}|}
\hline
\textbf{Method} & \textbf{Correlation} & \textbf{\(R2\) Score} & \textbf{\(NRMSE\)} &  \textbf{Average Error}\\
\hline
Color-Based & 0.82 (p $<$ \(10^{-7}\)) & 0.67  & 0.10 & 8.62 $\pm$ 0.79 \% \\
\hline
Gray-Based & 0.85 (p $<$ \(10^{-7}\)) & 0.73 & 0.09 & 7.89 $\pm$ 0.80 \% \\
\hline
Combination of color and gray-based approach & 0.94 (p $<$ \(10^{-7}\)) & 0.88 & 0.06 &  4.62 $\pm$ 0.62 \%\\
\hline
\end{tabular}
\label{comparison_results}
\end{table}

The results for the “combined”  method were better than when using only the “gray-based” method or only the “color-based” method, as shown in tables \ref{comparison_per_part_results} and \ref{comparison_results}, demonstrating that the two methods contribute synergistically to the evaluation of pupil size. 
Since the best results were achieved with the “combined” method, all the subsequent results were obtained using this approach.

As explained in the paragraph \textit{Testing the Luminosity Effect Prediction Model (LEPM) in
a normal laboratory conditions and with non-monochrome images}, we tested the LEPM and calibration procedure on non-monochrome, non-primary color images and in both a dark and a lit environment. The test was conducted with 10 participants using a video composed of 27 \\monochrome images and 46 non-monochrome emotionally neutral images with different colors and brightness.

\begin{figure}[!ht]
\centering
\includegraphics[width=0.5\textwidth]{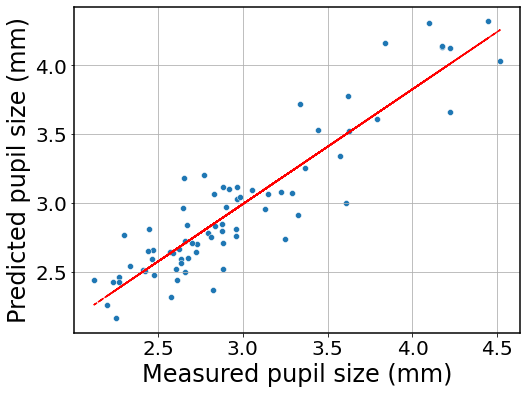}
\caption{Relationship between measured and predicted pupil size in a dark laboratory for Participant  \(IFL3\), with correlation: 0.91 (p \( < 10^{-7}\)), \(R2\): 0.83.}
\label{dark_corr_one_part}
\end{figure}

\begin{figure}[!ht]
\centering
\includegraphics[width=0.5\textwidth]{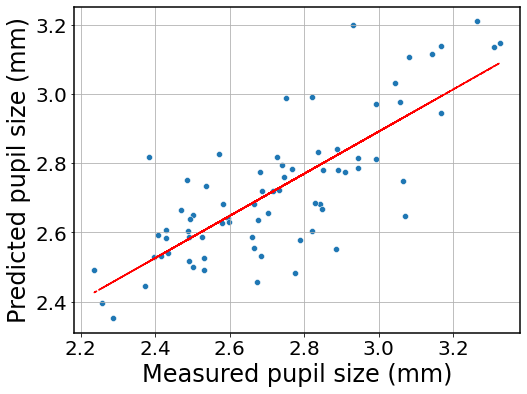}
\caption{Measured and predicted pupil size in a normal laboratory conditions for the Participant \(IFL3\), with correlation: 0.78 (p \(< 10^{-7}\)), \(R2\): 0.61.}
\label{light_corr_one_part}
\end{figure}

Figures \ref{dark_corr_one_part} and \ref{light_corr_one_part} show the results for participant IFL3 in a dark and a normal laboratory conditions, respectively. The average results across participants, obtained in the dark and normal laboratory conditions, are shown in Table \ref{comparison_results_unlit_lit_lab}. 

\begin{figure}[!ht]
\centering
\includegraphics[width=0.5\textwidth]{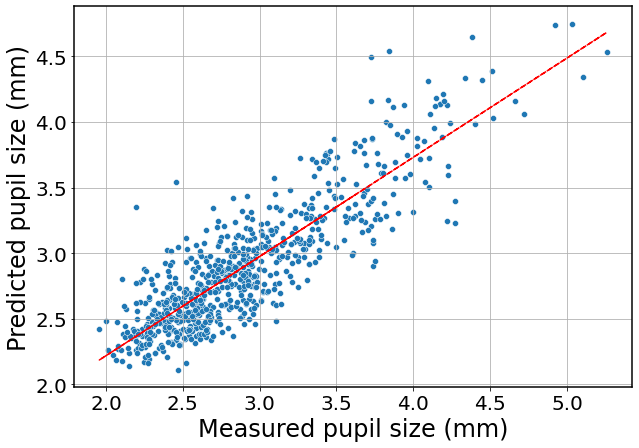}
\caption{Measured and predicted pupil size in a dark laboratory for all the participants, with correlation: 0.87 (p \( < 10^{-7}\)), \(R2\): 0.76.}
\label{dark_corr}
\end{figure}

\begin{figure}[!ht]
\centering
\includegraphics[width=0.5\textwidth]{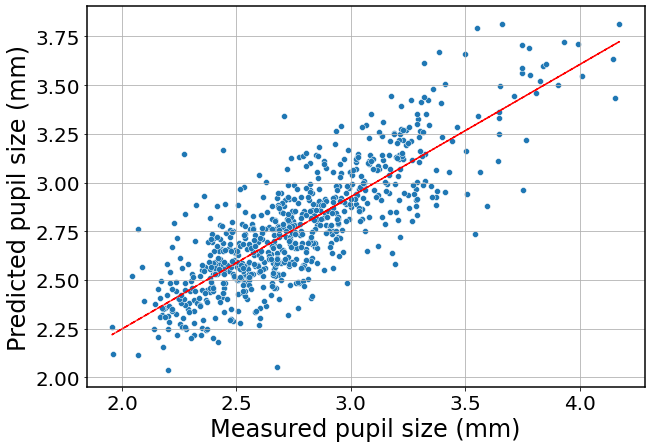}
\caption{Measured and predicted pupil size in a normal laboratory conditions for all the participants, with correlation: 0.82 (p \(< 10^{-7}\)), \(R2\): 0.68.}
\label{light_corr}
\end{figure}

Figures \ref{dark_corr} and \ref{light_corr} and Table \ref{comparison_results_unlit_lit_lab_aggregated}, instead, show the results obtained by aggregating the data from all the participants.

\begin{table}[!htbp]
\caption{\textit{Validation results of LEPM in dark light and normal laboratory conditions across all participants.}}
\centering
\begin{tabular}{|p{1.4cm}|p{1.7cm}|p{1cm}|p{1.2cm}|p{1.3cm}|}
\toprule
\textbf{Method} & \textbf{Correlation} & \textbf{\(R2\) Score} & \textbf{\(NRMSE\)} &  \textbf{Average Error}\\
\midrule
Dark light laboratory & 0.84 $\pm$ 0.061 (mean p $<$ \(10^{-7}\), max p $<$ \(10^{-7}\) & 0.70 $\pm$ 0.101  & 0.12 $\pm$ 0.020 & 7.58\% $\pm$ 1.61\% \\
\midrule
normal laboratory conditions & 0.76 $\pm$ 0.045 (mean p $<$ \(10^{-7}\), max p $<$ \(10^{-7}\) & 0.58 $\pm$ 0.0680 & 0.14 $\pm$ 0.014 & 7.72\% $\pm$ 1.66\% \\
\bottomrule
\end{tabular}
\label{comparison_results_unlit_lit_lab}
\end{table}

\begin{table}[!htbp]
\caption{\textit{Validation results of LEPM in dark light and normal laboratory conditions by aggregating data from all participants.}}
\centering
\begin{tabular}{|p{1.4cm}|p{1.7cm}|p{1cm}|p{1.2cm}|p{1.3cm}|}
\toprule
\textbf{Method} & \textbf{Correlation} & \textbf{\(R2\) Score} & \textbf{\(NRMSE\)} &  \textbf{Average Error}\\
\midrule
Dark light laboratory & 0.87 (mean p $<$ \(10^{-7}\), max p $<$ \(10^{-7}\) & 0.76 & 0.09 & 7.28\% $\pm$ 0.82\% \\
\midrule
normal laboratory conditions & 0.82 (mean p $<$ \(10^{-7}\), max p $<$ \(10^{-7}\) & 0.68 & 0.11 & 7.50\% $\pm$ 0.83\% \\
\bottomrule
\end{tabular}
\label{comparison_results_unlit_lit_lab_aggregated}
\end{table}

The average error between the two conditions is about 0.2\%, and the difference in \(NRMSE\) is about 0.02, highlighting the model's effectiveness across dark and normal laboratory conditions settings. Additionally, the model performs consistently across various monitor brightness and contrast settings and under different ambient lighting levels. Therefore, there was no need to modify the model to account for changes in environmental luminosity, given that our calibration procedure takes this into account effectively. However, we repeated the calibration every 20 minutes for very long experimental sessions.

\subsection{Results of the Arousal Detection Model (ADM)}

\begin{figure*}[!h]
\centering
\subfloat[High-Arousal]{\includegraphics[width=3.3in]{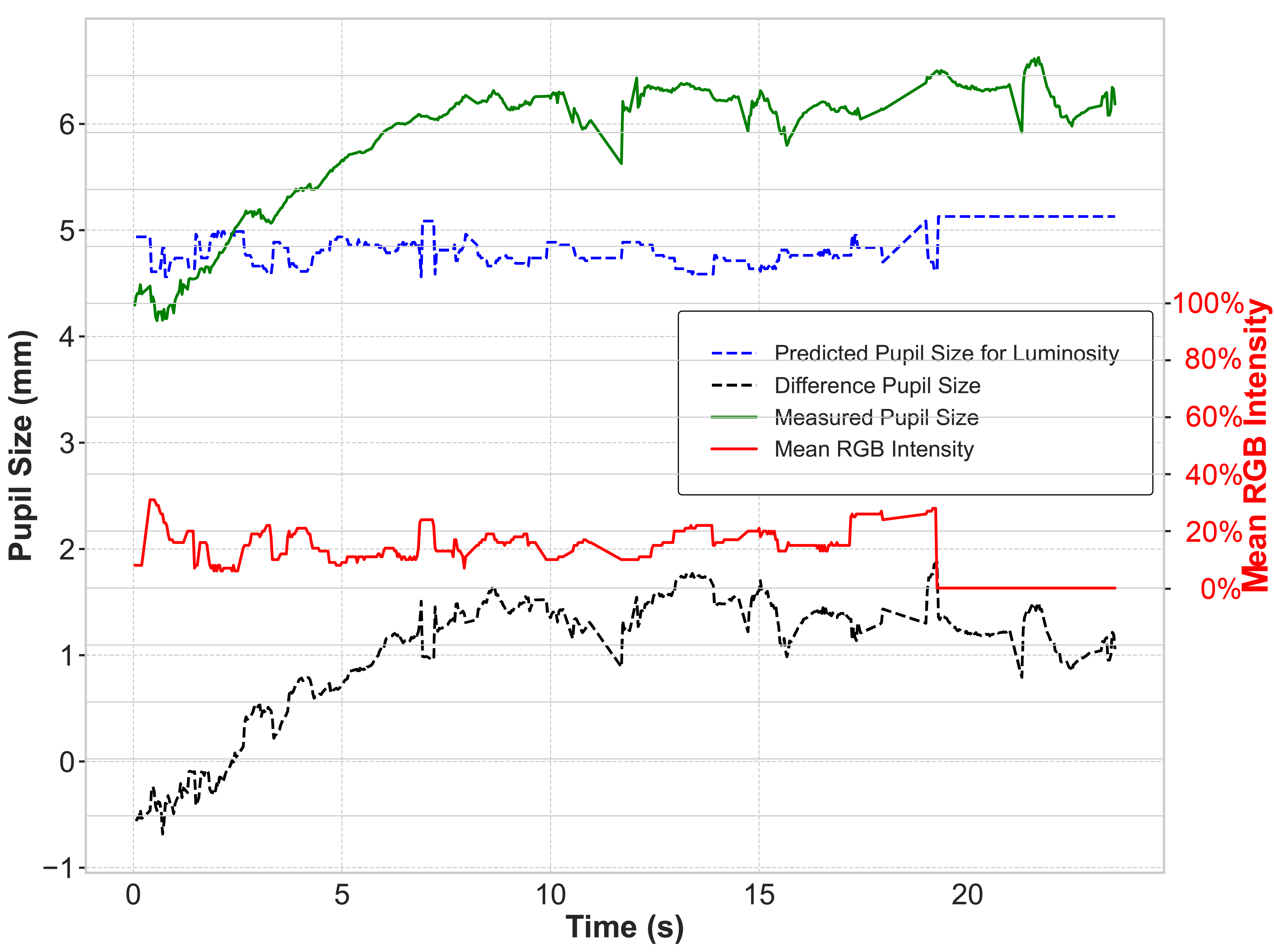}%
\label{HA}}
\hfill
\subfloat[Low-Arousal]{\includegraphics[width=3.3in]{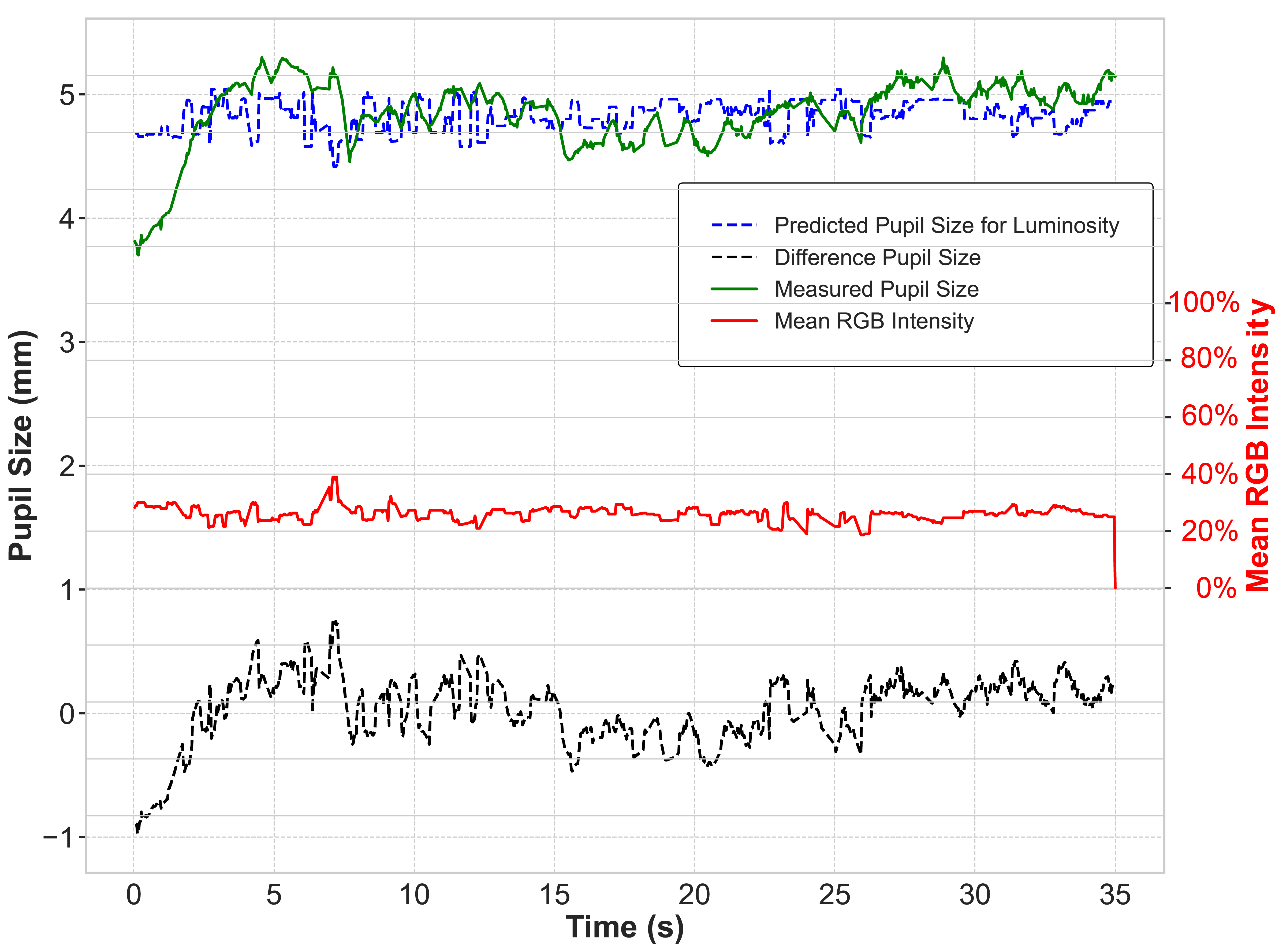}%
\label{LA}}
\caption{Plot of participant XPI3pA's response for selected high-arousal (a) and low-arousal (b) videos, showing measured pupil size (green), predicted pupil size (blue), average RGB intensity (red), and arousal-induced pupil size (black).}
\label{fig_sim}
\end{figure*}

To test the ADM, we predicted the arousal of participants watching video clips and compared it with the ground truth (actual pupil size). 
For each participant and each video clip, we plotted the value of the measured pupil size \(\PS_\text{measured}\) throughout the video clip.  This is shown in Figure \ref{fig_sim} (green line) for a particular participant, and a video with increasing high emotional intensity (Figure \ref{fig_sim}(a), and one with low intensity (Figure \ref{fig_sim}(b)). In the same figure, we show the average RGB value frame by frame, representing the luminous intensity of the video (red line). We then calculated the pupil size component due to luminosity \(\PS_\text{luminosity}\) using the LEPM (see Figure \ref{fig_sim}, blue line). Finally, we calculated the pupil size component due to arousal \(\PS_\text{arousal}\) by subtracting \(\PS_\text{luminosity}\) from \(\PS_\text{measured}\), according to Equation \ref{arousal_eq} (see the black line in Figure \ref{fig_sim}).

Our analysis revealed a clear distinction in the arousal responses between high-arousal and low-arousal video clips, as shown in the example illustrated in Figure \ref{fig_sim}. In fact, in the left panel (high arousal), the pupil size component due to arousal \(\PS_\text{arousal}\) (black line) assumes higher values than in the right panel (low arousal). 

\begin{figure}[!ht]
\centering
\includegraphics[width=0.5\textwidth]{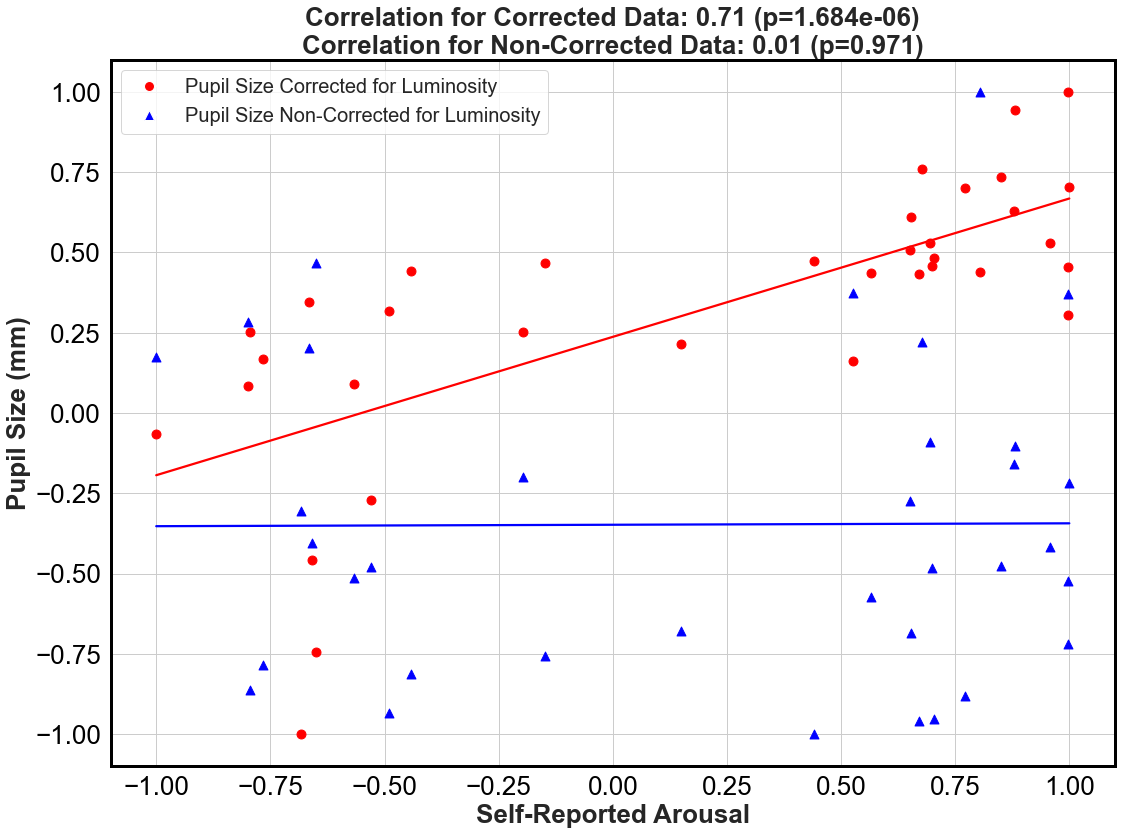}
\caption{Pupil size comparison with and without luminosity correction versus self-reported arousal for Participant XPI3pA. The red circles represent pupil size corrected for luminosity with the red linear regression (LR) line, while the blue triangles indicate non-corrected pupil size with the blue LR line.}
\label{part_corr_plot}
\end{figure}

We then calculated the average of the \(\PS_\text{arousal}\) (black line in Figure \ref{fig_sim}) in the salient intervals, as explained in the paragraph \textit{Model Testing} of section \ref{Arousal detection Model (ADM) Development}. For example, for the videos represented in Figure \ref{fig_sim}, the salient intervals were [0s,~20s] for video clip (a) and [0s,~10s] for video clip (b). Finally, we plotted the pupil size corrected for luminosity \(\PS_\text{arousal}\) versus the self-reported arousal \(\text{Arousal}_\text{self-reported}\) for all video clips and each participant, which is shown in Figure \ref{part_corr_plot} for participant XPi3pA, where the red circles correspond to each video clip (see figure \ref{part_corr_plot} for more information). For that participant, we obtained a correlation of 0.71 ($p = 1.6\cdot 10^{-6}$;  see the red line). For the measured pupil size, non-corrected for luminosity \(\PS_\text{measured}\), the correlation with the ground truth arousal was much worse -- see participant XPi3pA in Figure \ref{part_corr_plot}, where the blue dots correspond to each video. clip. We obtained a correlation of 0.01 (p = 0.971) for that participant (see blue line in the figure \ref{part_corr_plot}).

Table \ref{tab:corr_metrix} shows the average of all participants. Correcting pupil size for luminosity dramatically increased correlation compared to non-correcting it. The most surprising result, however, was that \textit{pupil size had no predictive power for arousal \underline{without} correcting for luminosity}, as shown by the correlation with self-reported arousal not being significantly different from zero (mean p = 0.2283).

\begin{figure}[!ht]
\centering
\includegraphics[width=0.5\textwidth]{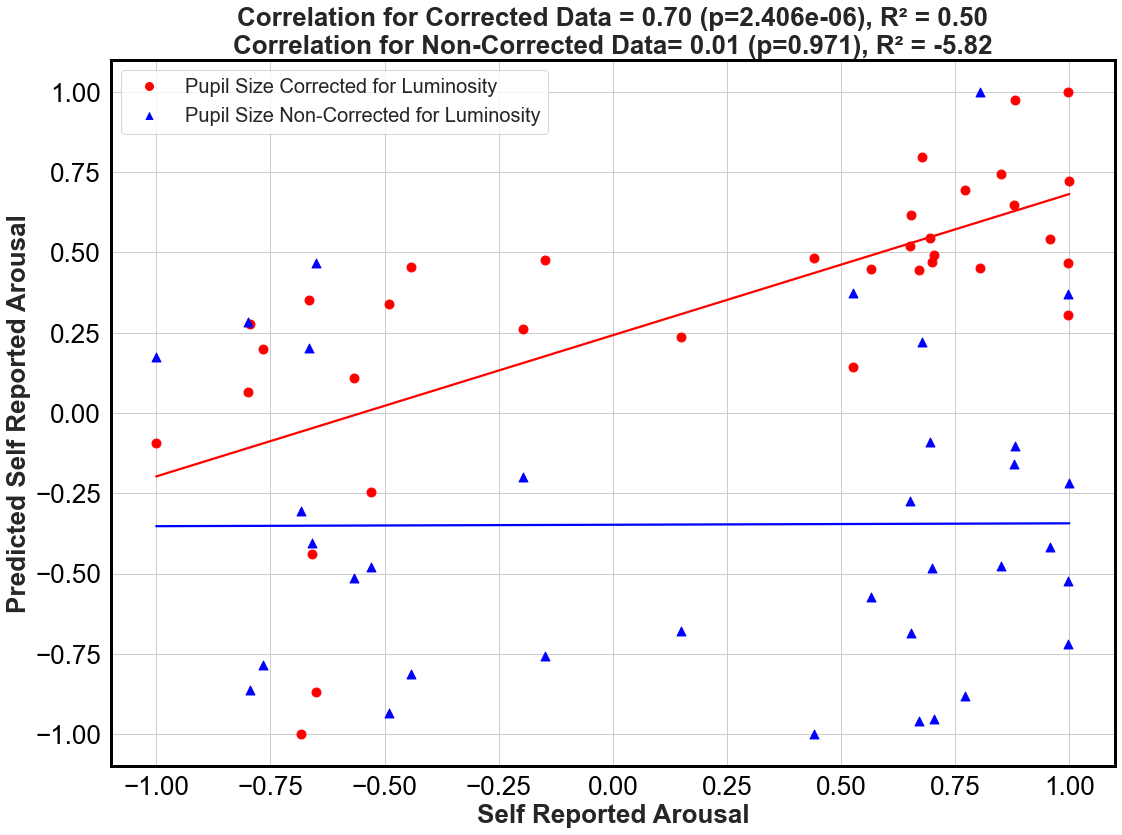}
\caption{Predicted arousal versus self-reported arousal with and without the use of correction for the luminosity for Participant XPI3pA.}
\label{fig_corr_participant_pred_arousal}
\end{figure}

\begin{table}[!htbp]
    
    \caption{Relationship between Self-Reported Arousal and Pupil Size with and without Correction for Luminosity} 
    \centering
    \begin{tabular}{ |p{1.4cm}|p{3cm}|p{3cm}| }
        \toprule
        \textbf{Metrics} & \textbf{Corrected for Luminosity} & \textbf{Non-Corrected for Luminosity} \\
        \midrule
        Correlation & 0.65 $\pm$ 0.106 (mean p =  0.0025, max p =  0.096) & 0.26 $\pm$ 0.150 (mean p =  0.2283, max p =  0.971) \\
        \midrule
        \(NRMSE\) & 0.27 $\pm$ 0.036 & 0.42 $\pm$ 0.054 \\
        \midrule
        $R2$ & 0.436 $\pm$ 0.125 & 0.09 $\pm$ 0.089 \\
        \bottomrule
    \end{tabular}
    \label{tab:corr_metrix}
\end{table}
To further test the model performance, we computed the predicted \(\widehat{\text{Arousal}}_{\text{self-reported}}\) utilizing the equations \ref{arousal_ps_lin_eq} and \ref{ps_arousal_inverted_eq} and using a leave-one-participant out cross-validation as explained in the paragraph \textit{ADM Testing}. We plotted the predicted arousal against the ground truth one \(\text{Arousal}_\text{self-reported}\) both with and without correction for luminosity, as shown in Figure \ref{fig_corr_participant_pred_arousal} for participant XPI3pA.

The average results across all participants are shown in the left panel in Table \ref{tab:indscal_vs_fa}. Again, correcting for luminosity produces a dramatic increase in correlation compared to uncorrected pupil size. While the left panel shows the results obtained by calculating self-reported arousal using individual scaling, the right panel shows the results obtained by calculating self-reported arousal using factor analysis. As mentioned in the paragraph \textit{Development of Arousal Detection Model}, INDSCAL works much better than factor analysis in this case.

\begin{table*}[!htbp]
    \centering
    \caption{Relationship between predicted and self-reported arousal with and without correction for luminosity using INSCAL and Factor analysis.}
    \label{tab:indscal_vs_fa}
    \resizebox{\textwidth}{!}{ 
    \begin{tabular}{ |>{\raggedright\arraybackslash}p{2cm}|p{3.5cm}|p{3.5cm}||>{\raggedright\arraybackslash}p{2cm}|p{3.5cm}|p{3.5cm}| }
        \toprule
        \multicolumn{3}{|c||}{\textbf{INDSCAL}} & \multicolumn{3}{c|}{\textbf{Factor Analysis}} \\
        \midrule
        \textbf{Metrics} & \textbf{Corrected for Luminosity} & \textbf{Non-Corrected for Luminosity} & \textbf{Metrics} & \textbf{Corrected for Luminosity} & \textbf{Non-Corrected for Luminosity} \\
        \midrule
        Correlation & 0.65 $\pm$ 0.12 (mean p =  0.0025, max p =  0.096) & 0.26 $\pm$ 0.15 (mean p =  0.2283, max p =  0.971) & Correlation & 0.33 $\pm$ 0.12 (mean p =  0.1397, max p =  0.367) & 0.11 $\pm$ 0.15 (mean p =  0.3466, max p 0.567) \\
        \midrule
        $R2$ & 0.43 $\pm$ 0.12 & 0.09 $\pm$ 0.089 & $R2$ & 0.11 $\pm$ 0.12 & 0.07 $\pm$ 0.09 \\
        \midrule
        \(NRMSE\) & 0.27 $\pm$ 0.036 & 0.42 $\pm$ 0.054 & \(NRMSE\) & 1.12 $\pm$ 0.73 & 1.75 $\pm$ 0.95 \\
        \bottomrule
    \end{tabular}
    }
\end{table*}

Finally, we noticed that, in the case of pupil size corrected for luminosity, the average values of the coefficients of Equation \ref{arousal_ps_lin_eq} were \(a\) = 0.3463 $\pm$ 0.0551, \(b\) = -0.0126 $\pm$ 0.00038 and the fit was quite good (\(R2\) = 0.567 $\pm$ 0.073). The variance of the coefficients and \(R2\) arises because leaving one participant out at a time changes both the slope and the intercept of the regression line. However, this variation is minuscule (the standard deviation is much smaller than the mean), indicating that our sample was sufficiently large.
When pupil size was not corrected for luminosity, the average values of the coefficients were \(a\) = 0.36723 $\pm$ 0.06341, \(b\) = 3.8296 $\pm$ 0.0120, and the fit was inferior (\(R2\) = 0.023 $\pm$ 0.007).
 
\subsection{Testing our model with ground truth obtained from independent judges rather than being self-reported} \label{model_without_self_reported_arousal}

We investigated how important it is to use self-reported arousal as ground truth. To do this, we used the approach other researchers have used: asking 10 independent judges (new participants) to give an arousal value to each video (e.g., Raiturkar et al. ~\cite{raiturkar2016decoupling}). For each clip, we took the average value. We then used that arousal value in our method, instead of the self-reported one, to observe how much the average arousal of the independent judges could be predicted by pupil size corrected for the luminosity of our 47 participants. The comparison between the results obtained with our model using as arousal values those self-reported by the 47 participants, i.e., what we described so far, and the results obtained with our model using as arousal values those given by the 10 independent judges, is shown in the last column of the Table \ref{tab:comparison_matrix_with_other_studies}. The model trained on the arousal reported by the 10 independent judges provides clearly worse results than the one trained on the arousal reported by the 47 participants. For example, there is a worsening in \(NRMSE\) of 188\%. This is because when training a model, one must use a ground truth recorded from the same subjects from which the predictors are recorded, in this case, respectively, the self-reported arousal and pupil size of the 47 participants in the study. The reason is that the biosignals recorded from a given participant convey information about that participant's emotional state.

\begin{table*}[!htbp]
    \caption{Comparison of the relationship between predicted and self-reported arousal in our model versus others' models.} 
    \centering
    \begin{tabular}{|p{1.5cm}|p{3cm}|p{3cm}|p{3cm}|p{3cm}|}
        \toprule
        \textbf{Metric} & \textbf{Our Model} & \textbf{Hyperbolic Model~\cite{nakayama2021controlling}} & \textbf{Linear Model~\cite{raiturkar2016decoupling, asano2021neural}} & \textbf{Our Model (without self-reported arousal)~\cite{raiturkar2016decoupling}} \\
        \midrule
        Correlation & 0.65 $\pm$ 0.106 (mean p =  0.0025, max p =  0.096) & 0.26 $\pm$ 0.150 (mean p =  0.1953, max p =  0.9566) & 0.26 $\pm$ 0.146 (mean p =  0.2260, max p =  0.9892) & 0.38 $\pm$ 0.074 (mean p =  0.0346, max p =  0.2886) \\
        \midrule
        \(NRMSE\) & 0.27 $\pm$ 0.036 & 0.41 $\pm$ 0.055 & 0.42 $\pm$ 0.052 & 0.78 $\pm$ 0.283 \\
        \midrule
        $R2$ & 0.436 $\pm$ 0.125 & 0.10 $\pm$ 0.087 & 0.07 $\pm$ 0.086 & 0.153 $\pm$ 0.054 \\
        \bottomrule
    \end{tabular}
    \label{tab:comparison_matrix_with_other_studies}
\end{table*}

\subsection{Results of the Gradient Boost Regression-based Self-reported Arousal Prediction Model (GBR-SAPM)}

This section presents the results obtained using the GBR-SAPM model. Again, taking into account the effect of luminosity on pupil size had a significant effect. In fact, correcting pupil size for luminosity significantly improved the model's performance, as shown in Table~\ref{tab:pupil_correction_comparison}. The model trained on corrected features achieved an \(R2\) of 0.556 $\pm$ 0.085, an \(NRMSE\) of 0.229 $\pm$ 0.022, and a statistically significant Pearson correlation of 0.765 $\pm$ 0.047 (mean p $<$ \(10^{-7}\), max p $<$ \(10^{-7}\)), across all participants. These results suggest that the corrected pupil size features provided a reliable basis for predicting arousal. In contrast, the model trained on non-corrected features showed much weaker performance across all participants: an \(R2\) of 0.235 $\pm$ 0.097, an \(NRMSE\) of 0.301 $\pm$ 0.020, and a non-significant Pearson correlation of 0.521 $\pm$ 0.116 (mean p = 0.345, max p = 0.863), emphasizing the degradation in model accuracy when luminosity is not accounted for.

\begin{table*}[h!]
\centering
\caption{Comparison of Emotion Prediction Performance Using Corrected vs. Non-Corrected Pupil Size Features}
\label{tab:pupil_correction_comparison}
\begin{tabular}{|p{2cm}|p{2cm}|p{2cm}|p{2cm}|p{7cm}|}
\toprule
\textbf{Correction} & \textbf{Target} & \textbf{ \(R2\) Score} & \textbf{\(NRMSE\)} & \textbf{Pearson $r$ (p )} \\
\midrule
\multirow{2}{*}{Corrected} 
    & Arousal & 0.556 $\pm$ 0.085 & 0.229 $\pm$ 0.022 & 0.765 $\pm$ 0.047 (mean p $<$ $10^{-7}$,max p $<$ $10^{-7}$) \\
    & Valence & 0.259 $\pm$ 0.251 & 0.295 $\pm$ 0.041 & 0.595 $\pm$ 0.082 (mean p = 0.0012, max p = 0.075) \\
\midrule
\multirow{2}{*}{Non-Corrected} 
    & Arousal & 0.235 $\pm$ 0.097 & 0.301 $\pm$ 0.020 & 0.521 $\pm$ 0.116 (mean p = 0.345, max p = 0.863) \\
    & Valence & 0.205 $\pm$ 0.242 & 0.306 $\pm$ 0.039 & 0.566 $\pm$ 0.065 (mean p = 0.0020, max p = 0.013) \\
\bottomrule
\end{tabular}
\end{table*}

For valence, as expected, the results were worse and the benefit of the luminosity correction was less pronounced. The corrected model achieved an \(R2\) of 0.259 $\pm$ 0.251 and a Pearson correlation of 0.595 $\pm$ 0.082 (mean p = 0.0012, max p = 0.075) compared to 0.205 $\pm$ 0.242 and 0.566 $\pm$ 0.065 (mean p = 0.0020, max p = 0.013) for the non-corrected model, across all participants. The \(NRMSE\) also slightly improved from 0.306 $\pm$ 0.039 to 0.295 $\pm$ 0.041, as presented in Table~\ref{tab:pupil_correction_comparison}.
\begin{figure}[!t]
    \centering
    \includegraphics[width=0.5\textwidth]{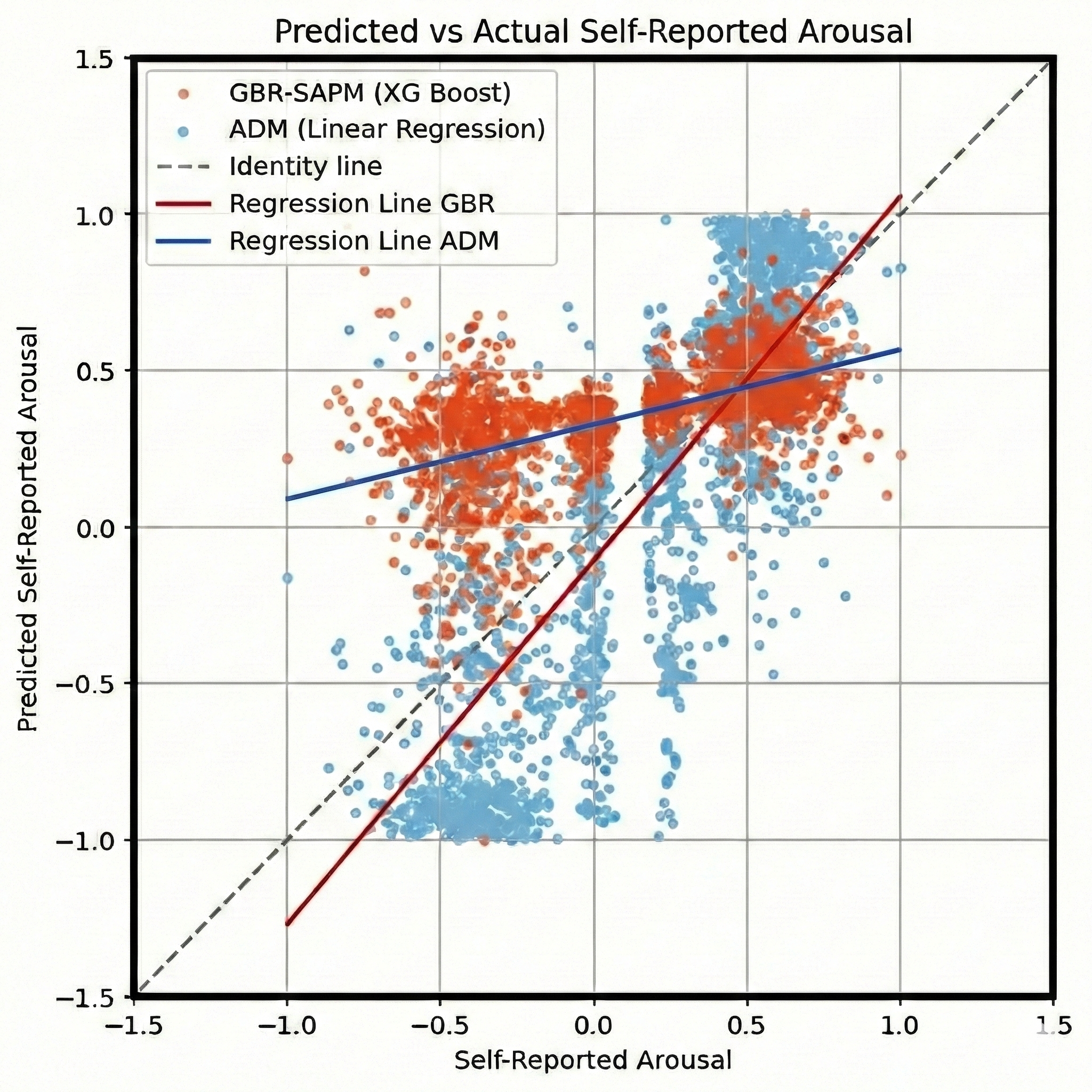}
    \caption{Predicted vs. Self-Reported Arousal graph to compare ADM (Linear Regression) and GBR-SAPM (XGBoost)}
    \label{fig:predicted_vs_actual}
\end{figure}

We can conclude that performance improves when moving from a linear regression model (ADM) to a more complex one such as XGBoost (GBR-SAPM). In fact, when the brightness effect is corrected, \(R2\) goes from 0.43 $\pm$ 0.12 to 0.556 $\pm$ 0.085, NMRSE goes from 0.27 $\pm$ 0.036 to 0.229 $\pm$ 0.022 and Pearson's correlation goes from 0.65 $\pm$ 0.12 (mean p = 0.0025) to 0.765 $\pm$ 0.047 (mean p $<$ $10^{-7}$).  

In Figure \ref{fig:predicted_vs_actual}, we have reported the predicted arousal values compared to the self-reported ones, for both the ADM and GBR-SAPM models and for all participants. The GBR-SAPM model has greater predictive power than the ADM model, and the slope of the regression line of predicted arousal points vs. self-reported arousal goes from 0.22 for the ADM model to 0.89 for the GBR-SAPM model. One reason why the GBR-SAPM model outperforms the ADM model may be that it does not assume a linear relationship between total pupil size, the part driven by emotional arousal, and the part driven by luminosity, which is not true, as recently discovered\cite{pan2022arousal}.

\section{Discussion}
The aim of this study was to develop and validate a scalable, non-linear computational method for separating the Pupillary Light Response (PLR) from emotional arousal during the viewing of naturalistic videos. Our results demonstrate that the isolated arousal-sensitive residual provides a significantly more accurate representation of subjective emotional states compared to raw pupil diameter. A secondary aim was to create a realistic tool for researchers to be used in typical lab conditions, without the need for specialized equipment.

\subsection{Physiological Validity and the Psycho-sensory Residual}
Our results provide strong empirical support for the \enquote{final common pathway} theory~\cite{mathot2018pupillometry, strauch2022pupillometry}, which proposes that the pupil acts as an integrated readout where independent neural inputs from both the light reflex and psychological responses are combined~\cite{mathot2018pupillometry, strauch2022pupillometry}. By effectively suppressing PLR through our calibrated exponential model, we have isolated the underlying autonomic activation, referred to here as the PPR component. The superior performance of both the simple linear regression-based model (ADM) and the more complex model (GBR-SAPM) when using the corrected residual compared to the raw signal suggests that \\luminosity-driven fluctuations act as non-linear noise that obscures affective information. Our results do not constitute a direct verification of the \enquote{final common pathway} theory, but do demonstrate that, once the variance related to luminosity is eliminated, pupil responses show a stronger and more systematic relationship with subjective self-reported arousal, as would be expected if autonomic activation reflected a convergent pathway. This validation aligns with the decoupling approach proposed by Raiturkar et al.~\cite{raiturkar2016decoupling}, whilst also providing a more precise physiological fit through non-linear modeling rather than linear approximation. Furthermore, the technical rigor of our exponential fit is supported by advanced optical studies, such as the Spatially Weighted Corneal Flux Density approach by Zhang et al.~\cite{zhang2019pupil}, even though our model simplifies these principles for laboratory usability. 

\subsection{Two-Stage Modeling}
A goal of this research was to ensure that the luminosity-corrected signal is immediately usable for affective inference without requiring exhaustive post-processing. To this end, we developed two distinct modeling stages. The first stage, the Arousal Detection Model (ADM), utilizes a simple linear regression to relate the corrected pupil residual to arousal. This was a deliberate design choice to demonstrate that once the complex non-linear effects of luminosity are removed, even a straightforward linear approach yields significant predictive improvements. This makes our methodology highly accessible to researchers in psychology and neurobiology who may not have specialized expertise in advanced machine learning. In the second stage, we implemented the more sophisticated Gradient Boosted Regression-based Self-reported Arousal Prediction Model (GBR-SAPM). We achieved superior performance metrics, illustrating that, although the corrected signal is robust in its raw form, advanced machine learning can further resolve intricate physiological dynamics. 

\subsection{Multiplicative Light-Arousal Interaction and Limit of the Additive Model}
A critical finding in this work is that while linear models often perform well under limited luminosity variations, they fail to capture the inherently non-linear and multiplicative interaction between arousal and light. As demonstrated by Pan et al.~\cite{pan2022arousal}, emotional arousal effects are not monolithic and interact differently depending on background luminance, with strongest effects occurring at very low luminances. Our model accounts for this non-linear structure and provides a more precise physiological fit than linear approximations. This distinction is vital for laboratory realism, where screen-based stimuli vary unpredictably. By utilizing an exponential fit (LEPM) and a non-linear regressor (GBR-SAPM), we move beyond the \enquote{additive} assumption that characterizes simpler decoupling frameworks.

\subsection{Methodological Scalability and Ecological Validity}
A significant innovation of this work is its scalability for typical psychological research. Traditional experimental designs require the use of controlled visual stimuli and isoluminant conditions to achieve stable and uniform luminosity across stimuli, an artificial process that is laborious and difficult to adapt to videos. Our method eliminates this requirement with a one-time individual calibration procedure that accounts for inter-subject differences and monitor configurations~\cite{pamplona2009photorealistic}. We have also taken into account the fluctuating nature of ambient light, a critical factor that is often overlooked in controlled laboratory studies. Our work will allow researchers to utilize diverse video content without specialized hardware like photometers, bridging the gap between high-precision engineering models and the practical operational needs of behavioral science.

\subsection{Holistic Experience vs. Dual-Task Confound}
Our research focused specifically on the holistic emotional response to short video clips. We intentionally used brief stimuli to evoke a single, consistent emotional state, and our ground truth was obtained through a 
single post-stimulus self-report. Our stimuli were designed to elicit a single, relatively stable emotional state per clip, and because clip durations (at least 8 seconds) exceeded the typical latency of pupillary responses, we considered the average value of pupil size over the entire video rather than transient dynamics. This choice prioritizes robustness and interpretability for holistic emotional states, though we acknowledge that dynamic features may carry additional information in paradigms involving rapid affective fluctuations. Although greater temporal granularity is a common computational goal, requiring participants to provide continuous, second-by-second self-reports can introduce dual-task interference. In such scenarios, the cognitive effort of monitoring and reporting one's internal state can compromise the authenticity of the emotional experience.  However, there are situations, such as when playing a video game, where it makes sense to provide continuous feedback on one's state of arousal, using special devices, like RankTrace~\cite{lopes2017ranktrace}.

\subsection{Validation Strategy: Self-Report vs. Independent Judges}
In this study, we used a relatively large sample, compared to existing research, with over 40 participants, a collection of over 30 emotional video clips, and self-reported arousal ratings for each participant and each video. To our knowledge, no previous research on pupil size measurement has incorporated self-reported arousal ratings with this level of detail. Some authors have hypothesized and described in their articles the emotional content of the videos shown to participants without validation from the participants themselves~\cite{tarnowski2020eye},~\cite{nakayama2021controlling},~\cite{asano2021neural, asano2021pupil}. Raiturkar and colleagues~\cite{raiturkar2016decoupling} used a less subjective approach based on the assessments of three independent judges and used them as ground truth, generating “arousal scores” for emotionally engaging scenes. A key finding in our study, however, was the need to use participants' self-assessments as the ground truth for interpreting biosignals. Indeed, when we used the ratings of 10 independent judges as a counter-control, the predictive accuracy of the model deteriorated significantly, with a 188\% increase in \(NRMSE\). This highlights that the pupillary response is an idiosyncratic physiological indicator, influenced by individual subjective experiences and cognitive biases. Although independent judges may be useful for the normative classification of stimuli, self-assessment is scientifically more consistent for training machines to interpret an individual's specific emotional state.  Our study shows that physiological responses, conscious emotional perception and their combination vary significantly between individuals. In our study, this is evident because the arousal and valence reported for a given video clip varied greatly between participants, underscoring the need to account for individual differences in studies of emotional perception.

\subsection{Practical Applications and Accessibility}
Thanks to its ease of implementation, our method has important practical applications, particularly in mental health monitoring and neuro-marketing. It requires only standard software for analyzing human perceptions, such as iMotions, and basic hardware for eye tracking~\cite{imotionsPupillometry101}, making it a cost-effective solution for myriad research contexts. Our method allows for the detection of emotional arousal without the need to analyze more complex or invasive biosignals, such as GSR, EEG, or ECG. In the field of mental health, where there is a persistent shortage of qualified counsellors, this approach offers an accessible, AI-based solution for emotion detection that could benefit under-served populations~\cite{jonnalagadda2023ensemble}. In consumer research, pupillometry has been used to assess emotional responses to advertisements but has not seen much popularity among marketers because of its low reliability~\cite{Wang2008Validity}. By eliminating luminosity effects, our model enables marketers to gauge the emotional arousal elicited by their advertisements much more accurately. Although extensive testing in real-world contexts is still needed to improve generalizability across different populations and stimuli, our method provides a solid foundation for more reliable affective assessment in these fields.

\subsection{Limitations and Future Directions}
In future work, we will include localized ocular factors in our model, such as gaze position or spatial light weighting, as discussed in the technical literature. Some of these factors, such as gaze position, are implicitly taken into account but not used as features.
A limitation of our work is that our models lack of granularity. In future work, we will make our model dynamic and increase temporal granularity without increasing the burden on participants. For this purpose, we want to use deep learning techniques, such as a neural network that incorporates temporal parameters, as demonstrated in~\cite{asano2021neural, asano2021pupil}.
While we believe the current GBR-SAPM model accounts for the multiplicative light-arousal interaction implicitly through its non-linear architecture, future developments will focus on accounting for this phenomenon explicitly to further refine the separation of signals.
Furthermore, while this article has focused exclusively on the pupil to isolate its predictive value, future work will integrate this signal with multimodal data, including heart rate and GSR, which were recorded during this study.
In a future study, we will also explore the extent to which pupil responses to emotional stimuli also vary depending on health conditions~\cite{grujic2024neurobehavioral}, personality traits~\cite{bradley2008pupil}, and emotional sensitivity~\cite{laurenzo2016pupillary}.

\section{Conclusion}
This research represents a significant advancement in the use of pupillometry for emotional arousal detection in naturalistic laboratory settings. By developing a model that effectively isolates the psychosensory component from the dominant pupillary light reflex, we have provided a solution to a long-standing methodological obstacle in affective computing. 

Our findings establish three critical conclusions: 

A) Correction is Mandatory: Raw pupil size lacks significant predictive power for measuring arousal during dynamic video viewing; accurate inference is only possible after removing the non-linear luminosity component. 

B) Subject-Centric Validation: For the purposes of training machines in which the target variable is the arousal experienced by participants, participants' self-reports provide a more appropriate reference than external judges' assessments, which reflect the normative properties of the stimulus rather than subjective experience. 

C) Methodological Scalability: The proposed calibrated exponential model offers a \enquote{realistic and scalable} alternative to complex engineering approaches, enabling high-fidelity research with standard eye-tracking hardware and diverse video stimuli.

While the current work focuses on holistic emotional responses to short video clips to preserve ecological validity and avoid dual-task reporting confounds, future research will extend this framework into a dynamic model. By integrating this corrected pupillary signal with other physiological measures such as GSR and heart rate, we aim to develop a robust, multi-modal system for real-time emotion detection across various domains, including healthcare and human-computer interaction.

\begin{acknowledgements}
We would like to thank Prof. Luca Ricolfi for inspiring the psychometric methods used, Akashdeep Nijjar, Sophia Carbonero and Dr Katerina Bourazeri for being an essential member of the research group, Dr Arno Onken and Cristina Del Prete for the precious advice, Kaylee Shurety, Flavia Popescu-Richardson and Ville Karhusaari for the continuous, kind and competent assistance, and Domiziana Falaschi for proofreading the manuscript.
\end{acknowledgements}


\printbibliography

\end{document}